\documentclass[twocolumn]{aastex62}
\usepackage{graphicx}	
\usepackage[caption=false]{subfig}
\usepackage{rotating}

\usepackage{amssymb}
\usepackage{color}
\usepackage{amsmath}
\usepackage{longtable}
\usepackage{threeparttablex}
\usepackage{lineno}

\usepackage{natbib}
\newcommand{\ie}{\textit{i.e.,}}
\newcommand{\eg}{\textit{e.g.,}}

\newcommand{\Dfour}{$D_n$(4000)}

\newcommand{\logM}{$\log(M_*/\rm{M}_\odot)$}
\newcommand{\logMdyn}{$\log(M_{\rm{dyn}}/\rm{M}_\odot)$}

\newcommand{\MDS}{$M_{\rm dyn}/M_*$}

\newcommand{\zspec}{$z_\textrm{spec}$}
\newcommand{\magazine}{MAGAZ3NE}

\newcommand{\OII}{\hbox{{\rm [O}\kern 0.1em{\sc ii}{\rm ]}}}
\newcommand{\OIIdub}{\hbox{{\rm [O}\kern 0.1em{\sc ii}{\rm ]$\lambda\lambda3726,3729$}}}
\newcommand{\OIII}{\hbox{{\rm [O}\kern 0.1em{\sc iii}{\rm ]}}}
\newcommand{\OIIIdub}{\hbox{{\rm [O}\kern 0.1em{\sc iii}{\rm ]$\lambda\lambda4959,5007$}}}
\newcommand{\OIIIfour}{\hbox{{\rm [O}\kern 0.1em{\sc iii}{\rm ]$\lambda4959$}}}
\newcommand{\OIIIfive}{\hbox{{\rm [O}\kern 0.1em{\sc iii}{\rm ]$\lambda5007$}}}

\newcommand{\Hbeta}{$\rm{H}\beta$}

\begin{document}

\title{\sc \magazine: High Stellar Velocity Dispersions for Ultra-Massive Quiescent Galaxies at $z\gtrsim3$} \footnote{The spectra presented herein were obtained at the W. M. Keck Observatory, which is operated as a scientific partnership among the California Institute of Technology, the University of California and the National Aeronautics and Space Administration. The Observatory was made possible by the generous financial support of the W. M. Keck Foundation.\\}
\shorttitle{Velocity Dispersions of UMGs}
\shortauthors{B. Forrest, et al.}

\correspondingauthor{Ben Forrest}
\email{bforrest@ucdavis.edu}

\author[0000-0001-6003-0541]{Ben Forrest}
	\affiliation{Department of Physics and Astronomy, University of California, Davis, One Shields Avenue, Davis, CA 95616, USA}
	\affiliation{Department of Physics and Astronomy, University of California, Riverside, 900 University Avenue, Riverside, CA 92521, USA}

\author[0000-0002-6572-7089]{Gillian Wilson}
	\affiliation{Department of Physics and Astronomy, University of California, Riverside, 900 University Avenue, Riverside, CA 92521, USA}

\author[0000-0002-9330-9108]{Adam Muzzin}
	\affiliation{Department of Physics and Astronomy, York University, 4700, Keele Street, Toronto, ON MJ3 1P3, Canada}
	
\author[0000-0001-9002-3502]{Danilo Marchesini}
	\affiliation{Department of Physics and Astronomy, Tufts University, 574 Boston Avenue, Medford, MA 02155, USA}

\author[0000-0003-1371-6019]{M. C. Cooper}
	\affiliation{Center for Cosmology, Department of Physics and Astronomy, University of California, Irvine,  4129 Frederick Reines Hall, Irvine, CA, USA}

\author[0000-0002-7248-1566]{Z. Cemile Marsan}
	\affiliation{Department of Physics and Astronomy, York University, 4700, Keele Street, Toronto, ON MJ3 1P3, Canada}

\author{Marianna Annunziatella}
	\affiliation{Department of Physics and Astronomy, Tufts University, 574 Boston Avenue, Medford, MA 02155, USA}
	\affiliation{Centro de Astrobiolog\'ia (CSIC-INTA), Ctra de Torrej\'on a Ajalvir, km 4, E-28850 Torrej\'on de Ardoz, Madrid, Spain}
	
\author[0000-0002-2446-8770]{Ian McConachie}
	\affiliation{Department of Physics and Astronomy, University of California, Riverside, 900 University Avenue, Riverside, CA 92521, USA}
	
\author{Kumail Zaidi}
\affiliation{Department of Physics and Astronomy, Tufts University, 574 Boston Avenue, Medford, MA 02155, USA}

\author{Percy Gomez}
	\affiliation{W.M. Keck Observatory, 65-1120 Mamalahoa Hwy., Kamuela, HI 96743, USA}

\author[0000-0001-8169-7249]{Stephanie M. Urbano Stawinski}
	\affiliation{Department of Physics and Astronomy, University of California, Irvine, CA 92697, USA}

\author{Wenjun Chang}
	\affiliation{Department of Physics and Astronomy, University of California, Riverside, 900 University Avenue, Riverside, CA 92521, USA}
	
\author{Gabriella de Lucia}
	\affiliation{INAF - Astronomical Observatory of Trieste, via G.B. Tiepolo 11, I-34143 Trieste, Italy}

\author[0000-0003-1181-6841]{Francesco La Barbera}
	\affiliation{INAF - Osservatorio Astronomico di Capodimonte, sal. Moiariello 16, 80131 Napoli, Italy}	
	
\author[0000-0003-2119-8151]{Lori Lubin}
	\affiliation{Department of Physics and Astronomy, University of California, Davis, One Shields Avenue, Davis, CA 95616, USA}
	
\author[0000-0002-7356-0629]{Julie Nantais}
	\affiliation{Departamento de Ciencias F\'isicas, Universidad Andres Bello, Fern\'andez Concha 700, Las Condes 7591538, Santiago, Regi\'on Metropolitana, Chile}

\author[0000-0002-0033-5041]{Theodore Pe\~na}	
	\affiliation{Department of Astronomy, University of Wisconsin-Madison, 475 N. Charter Street, Madison, WI 53706-1507, USA}

\author[0000-0003-3959-2595]{Paolo Saracco}	
	\affiliation{INAF - Osservatorio Astronomico di Brera, via Brera 28, 20121 Milano, Italy}
	
\author[0000-0001-7291-0087]{Jason Surace}	
	\affiliation{IPAC, Mail Code 100-22 Caltech 1200 E. California Blvd. Pasadena, CA 91125, USA}
	
\author[0000-0001-7768-5309]{Mauro Stefanon}
	\affiliation{Leiden Observatory, Leiden University, 2300 RA Leiden, The Netherlands}


\begin{abstract}

In this work we publish stellar velocity dispersions, sizes, and dynamical masses for 8 ultra-massive galaxies (UMGs; \logM$>11$, $z\gtrsim3$) from the Massive Ancient Galaxies At $z>3$ NEar-infrared (\magazine) Survey, more than doubling the number of such galaxies with velocity dispersion measurements at this epoch.
Using the deep Keck/MOSFIRE and Keck/NIRES spectroscopy of these objects in the $H$- and $K$-bandpasses, we obtain large velocity dispersions of $\sim400$ km s$^{-1}$ for most of the objects, which are some of the highest stellar velocity dispersions measured, and $\sim40$\% larger than those measured for galaxies of similar mass at $z\sim1.7$.
The sizes of these objects are also smaller by a factor of 1.5-3 compared to this same $z\sim1.7$ sample.
We combine these large velocity dispersions and small sizes to obtain dynamical masses.
The dynamical masses are similar to the stellar masses of these galaxies, consistent with a Chabrier initial mass function (IMF).
Considered alongside previous studies of massive quiescent galaxies across $0.2<z<4.0$, 
there is evidence for an evolution in the relation between the dynamical mass - stellar mass ratio and velocity dispersion as a function of redshift.
This implies an IMF with fewer low mass stars (\eg\ Chabrier IMF) for massive quiescent galaxies at higher redshifts in conflict with the bottom-heavy IMF (\eg\ Salpeter IMF) found in their likely $z\sim0$ descendants, though a number of alternative explanations such as a different dynamical structure or significant rotation are not ruled out.
Similar to data at lower redshifts, we see evidence for an increase of IMF normalization with velocity dispersion, though the $z\gtrsim3$ trend is steeper than that for $z\sim0.2$ early-type galaxies and offset to lower dynamical-to-stellar mass ratios.
\end{abstract}

\section{Introduction}

Deep near-infrared photometric surveys of the last decade have suggested larger numbers of massive galaxies at high redshifts than predicted by cosmological galaxy simulations \citep[\eg][]{Muzzin2013, Straatman2014, Sherman2019, Marsan2022}.
More recent simulations have better agreement with observations, but the discrepancy is still a factor of a few to ten at the highest masses \citep[though see ][]{Donnari2021,Lustig2022}.
In the last several years, spectroscopic confirmation of a handful of galaxies with stellar masses of \logM$>11$ and at redshifts of $z>3$ have shown that such galaxies do indeed exist in non-negligible numbers \citep{Marsan2015,Marsan2017,Glazebrook2017, Schreiber2018b, Tanaka2019, Valentino2020, Forrest2020a, Forrest2020b}, but a robust measurement of the number density of such galaxies is still lacking. 
This is largely due to the fact that the determination of stellar mass, particularly from photometry alone requires a number of assumptions which introduce the possibility for significant error.

There are numerous programs which determine galaxy parameters via spectral energy distribution (SED) fitting. 
Nearly all require some assumptions about the geometry of dust and dust extinction, the initial mass function (IMF) of star-formation, a parametric form of star-formation history, strength of emission lines, and choice of stellar population synthesis models, each of which play a role in the determined stellar mass of a galaxy.
For large populations of galaxies, the median mass determination appears sensitive to these choices with scatter  $\sim0.2$~dex \citep[\eg][]{Wuyts2009, Mobasher2015}, though the choice of code can lead to systematic offsets up to $0.3$~dex \citep{Muzzin2009, Leja2019}.
While these differences are perhaps tolerable, the differences for individual galaxies can greatly exceed these numbers in cases with significant flux contributions from strong emission lines \citep{Stark2013, Salmon2015, Forrest2017}, and active galactic nuclei \citep[AGN;][]{Leja2018}, as well as in outlier cases where photometric redshifts are highly discrepant from true redshifts, though this seems less common in massive galaxies even at $z>3$ \citep{Schreiber2018b, Forrest2020b}.

As a result, probing stellar masses independently of the above assumptions is valuable.
While the stellar velocity dispersion formally probes the total mass of a galaxy, the massive, high redshift galaxies of interest here typically have small sizes, and have central masses dominated by stars \citep[\eg][]{vanderWel2014,Straatman2015,Saracco2019}.
Locally, stellar velocity dispersion is well correlated with the luminosity and radius of elliptical galaxies \citep[\eg][]{Faber1976, Djorgovski1987, Dressler1987, Shu2012}, the mass of the central black hole \citep[\eg][]{Gebhardt2000, Kormendy2013}, mass-to-light ratio \citep[\eg][]{Cappellari2006}, and numerous other properties including galaxy color \citep{Wake2012} and stellar mass \citep[\eg][]{Zahid2016}.
Velocity dispersions have been studied out to higher redshifts as well, and many such correlations appear to hold for these data, though they may be offset from the local relations \citep[\eg][]{vanDokkum2009, Newman2010, Bezanson2012, Bezanson2013b, Thomas2013, vandeSande2013, Gargiulo2016, Hill2016, Belli2017}.
However, like the measurement of stellar masses, the measurement of stellar velocity dispersions holds the potential for systematic and statistical errors, the latter of which can of course be significant for low signal-to-noise (SNR) spectra.
The interpretation also requires careful analysis, as effects such as galaxy rotation and inclination can either increase or decrease measured velocity dispersions \citep{Bezanson2018, Newman2018b, Mendel2020}.

Still, stellar velocity dispersions can be used in concert with structural measurements to calculate dynamical masses, which are sensitive to the gravitational potential of a galaxy, and therefore to the contribution of dark matter as well as the contributions of dust, gas, and stars.
This then provides an effective upper limit on the stellar mass of a galaxy, independent of the numerous assumptions intrinsic to the calculation of stellar masses via SED fitting, including the shape of the initial mass function (IMF).

Variability in the IMF, which traces the number of stars formed as a function of their mass in a star-forming molecular cloud, can contribute to non-negligible differences in the determination of stellar mass as it sets the effective mass-to-light ratio.
The IMF of many galaxies, particularly local massive early-type galaxies (ETGs), is inferred via spectral fitting or
dynamical modeling to have a 'heavy' mass-to-light ratio (with respect to the MW distribution), such as that of the \citet{Salpeter1955} IMF,
which assumes a functional power-law with index $x=-2.35$ (termed `heavy' due to the larger effective mass-to-light ratio).
However, observations have suggested that the IMF is not universal \citep[see][for a review]{Hopkins2018a} and can vary over cosmic time, between galaxies, or as a function of galaxy radius, metallicity, stellar mass, or star formation density \citep[\eg][]{Cappellari2006, vanDokkum2008, Conroy2012, Cappellari2013a, Cappellari2013b, Kroupa2013, vanDokkum2017, Villaume2017, LaBarbera2019}.
As such it is important to note that any measurement of the IMF in a galaxy is a measurement of the super-position of the IMF during any and all episodes of star-formation in that galaxy.

Recently, \citet{Mendel2020} homogeneously analyzed 58 massive quiescent galaxies at $1.4<z<2.1$ and found that galaxies with higher stellar velocity dispersions at a given epoch prefer a heavier IMF such as that from \citet{Salpeter1955}, while galaxies with lower stellar velocity dispersions are better described by a lighter IMF such as the \citet{Chabrier2003} IMF.
This result agrees with lower redshift analysis from \citet{Posacki2015}, though the higher redshift galaxies have systematically higher velocity dispersions than lower redshift galaxies with the same dynamical-to-stellar mass ratio.

Measurements of velocity dispersion require spectra with reasonable signal-to-noise which are difficult to obtain for galaxies at earlier epochs.
As such, only six massive galaxies with stellar masses \logM$\gtrsim11$ at $z>3$ have measured stellar velocity dispersions \citep{Tanaka2019, Saracco2020, Esdaile2021}.
In this work we measure velocity dispersions for 8 additional massive galaxies at $z\gtrsim3$ using the MOSFIRE \citep{McLean2010, McLean2012} and NIRES \citep{JWilson2004} instruments on Keck, more than doubling the size of the current sample in the literature - 4/8 of these galaxies are more massive than any of the $z>3$ sample with velocity dispersions in the literature.
Combined with size measurements for these galaxies and values from the literature, we perform the first statistical comparison of dynamical and stellar masses at this early epoch using 14 massive galaxies.

We present the data in Section \ref{Sec:Data}, the velocity dispersion calculations and image analysis process in Section \ref{Sec:Analysis}, and then a discussion of the results in Section \ref{Sec:Res} and the main conclusions in Section \ref{Sec:Conc}.
All analysis here uses a $\Lambda$CDM cosmology with $H_0=70$ km s$^{-1}$ Mpc$^{-1}$, $\Omega_M=0.3$, and $\Omega_\Lambda=0.7$ as well as the AB magnitude system \citep{Oke1983}.
A \citet{Chabrier2003} IMF is used for calculation of stellar mass.

\section{Data} \label{Sec:Data}

\subsection{Parent Photometric Catalogs}

Targets selected for spectroscopic followup in the \magazine\ survey were drawn from parent photometric catalogs in the UltraVISTA DR1 \citep{Muzzin2013a}, UltraVISTA DR3 (Muzzin et al., in prep) and XMM-VIDEO (Annunziatella et al., in prep) fields.

The UltraVISTA survey \cite{McCracken2012} imaged over 1.62~deg$^2$ in the COSMOS field with  deep near-infrared \mbox{$Y$-,} \mbox{$J$-,} \mbox{$H$-,} and $K$-bandpasses.
The first data release \citep[DR1 catalogs;][]{Muzzin2013a} combined additional photometry from $0.15-24$~$\mu$m yielding a total of 30 bandpasses with 90\% completeness $K_s=23.4$ mag. 
Subsequent deep imaging over 0.84~deg$^2$ in the NIR furthered the value of the dataset, with DR3 \citep[][Muzzin et al., in prep]{Marsan2022} reaching deeper than DR1 by 1.1 mag in the $K_s$-band and $\sim1.2$ mag deeper in the IRAC $3.6$ and $4.5~\mu$m bandpasses \citep{Ashby2018}.
A total of up to 49 bandpasses in DR3 allowed for highly accurate galaxy spectral energy distributions (SEDs) and photometric redshift determinations, as well as detection of massive, quiescent galaxies at $z>3$ which are too faint for accurate characterization with optical photometry alone.

The VISTA Deep Extragalactic Observations \citep[VIDEO;][]{Jarvis2013} survey similarly acquired deep NIR imaging over several fields, including IRAC data from SERVS \citep{Mauduit2012} and the DeepDrill survey \citep{Lacy2021}.
Catalogs used in this work are built from VIDEO DR4 data over 5.1~deg$^2$ in the XMM-Newton Large Scale Structure (XMM) field with up to 22 bandpasses from $u$-band to IRAC 8.0 $\mu$m and a $5\sigma$ depth of $K_s=23.8$ mag (Annunziatella, et al., in prep).
While this catalog is somewhat shallower in $K$-band depth, it covers a wider area which is important for detection of the rare massive, quiescent objects at these redshifts.

\subsection{Near-Infrared Spectroscopy}

For this work we analyze $H$- and $K$-band spectroscopic observations from Keck-MOSFIRE \citep{McLean2010, McLean2012} taken as part of the \magazine\ survey \citep{Forrest2020b} and details of the spectroscopic target selection are provided therein.
The general survey observing strategy called for targeting ultra-massive galaxies (UMGs) in the $K$-band, where the strong emission features \OIIIfive\ and \Hbeta\ fall at the redshift of the sample.
On-the-fly reduction was used, and once a redshift was confirmed, observation of a UMG was stopped.
As such, UMGs with strong emission lines and only faint detection of the continuum have insufficient SNR to calculate a stellar velocity dispersion.

However, 6 of the 16 confirmed UMGs from \citet{Forrest2020b} have MOSFIRE observations in $H$-band where a greater number of spectral features lie (\eg\ \Dfour, Ca H\&K, and higher order Balmer absorption features), enabling a more reliable velocity dispersion calculation.
Since publication of \citet{Forrest2020b}, a redshift has also been obtained for 
an additional UMG, \mbox{COS-DR1-99209}, at z = 2.983 observed in both $H$- and $K$-band with MOSFIRE. 

For these seven galaxies, the MOSFIRE DRP was used to reduce the raw spectroscopy to 2D spectra.
From there, a custom code written by one of us (B.F.) was used to optimally extract a 1D spectrum and perform telluric corrections using stars observed on the same masks, modeled with the PHOENIX stellar models \citep{Husser2013}.
A more detailed description is provided in \citet{Forrest2020b}.
Galaxy photometry are fit in conjunction with a single-band spectrum using FAST++ \citep{Schreiber2018a} to obtain relative scaling of different spectral bandpasses.

In the case of XMM-VID1-2075, the only MAGAZ3NE UMG in this work without $H$-band MOSFIRE spectroscopy, $J$-, $H$- , and $K$-band NIRES spectroscopy was independently obtained (PI: Gomez; Gomez et al., in prep). 
A comparison of the two (very similar) $K$-band spectra is presented in Appendix~\ref{App:K2075}.
The NIRES data were reduced using Pypeit \citep[version 1.0.4;][]{Prochaska2020}. 
Pypeit flat fields the science data, performs wavelength calibration, models and subtracts the sky background, and performs a flux calibration.
A telluric correction was also calculated using Molecfit \citep{Smette2015, Kausch2015}.

In total, we thus present new velocity dispersions for 8 \magazine\ UMGs in this work: \mbox{COS-DR1-99209}, \mbox{COS-DR3-84674}, \mbox{COS-DR3-111740}, \mbox{COS-DR3-201999}, \mbox{COS-DR3-202019}, \mbox{XMM-VID1-2075}, \mbox{XMM-VID3-1120}, and \mbox{XMM-DR3-2457}.
We also include a ninth UMG from the MAGAZ3NE sample, \mbox{COS-DR3-160748}, which has a velocity dispersion from a high SNR spectrum taken with the LBT published in \citet{Saracco2020} as C1-23152.

\section{Analysis}\label{Sec:Analysis}

\subsection{Velocity Dispersions}

Absorption feature stellar velocity dispersions were calculated using the Penalized Pixel-Fitting method (pPXF) \citep{Cappellari2004,Cappellari2017} in conjunction with the UMG spectra.
This maintains consistency with the analysis of other $z>3$ massive galaxies \citep{Tanaka2019, Saracco2020, Esdaile2021}.


\begin{figure*}[tp]
	\centering{\includegraphics[width=\textwidth,trim=0in 0.25in 0in 0in, clip=true]{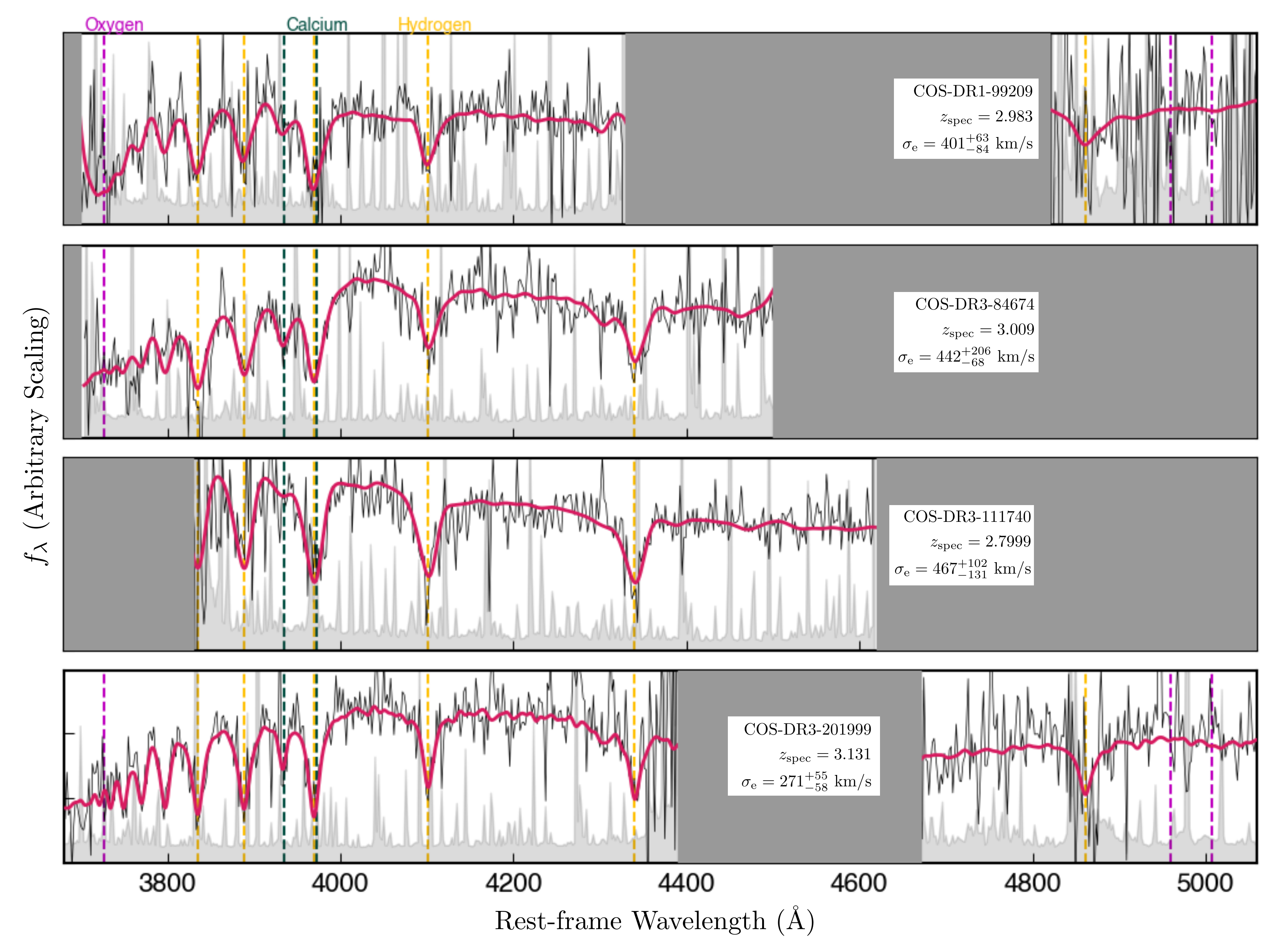}}
	\caption{The observed $H$- and $K$-band spectra for the MAGAZ3NE UMGs (black) and associated error spectra (gray). The best-fit pPXF model is shown in red. The wavelengths of prominent features from oxygen (magenta), calcium (green), and hydrogen (gold) are indicated as vertical dashed lines. When emission lines are present, we mask these features in the velocity dispersion fit.}
	\label{fig:fits1}
\end{figure*}
\begin{figure*}[tp]
        \ContinuedFloat
	\centering{\includegraphics[width=\textwidth,trim=0in 0.25in 0in 0in, clip=true]{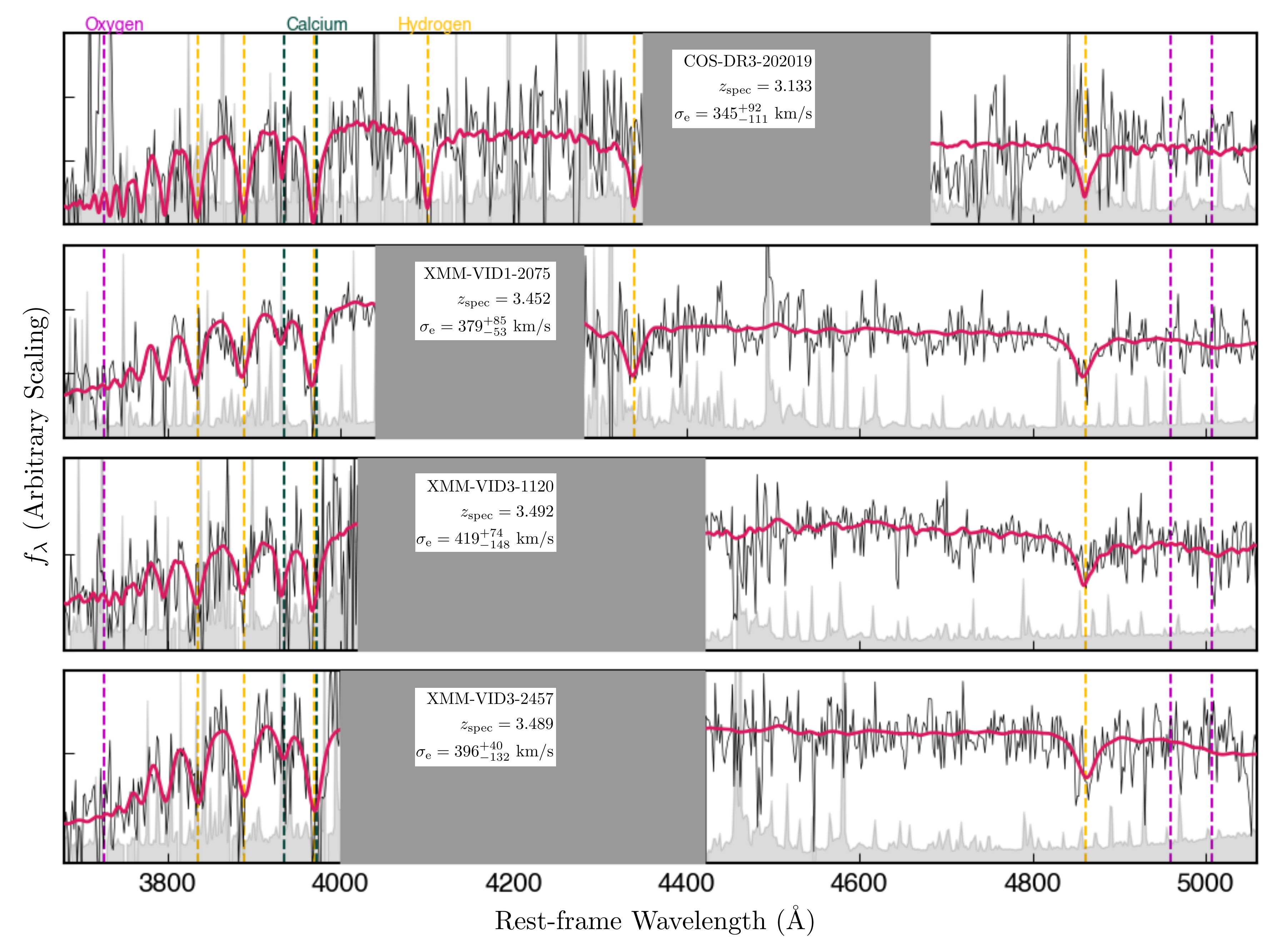}}
	\caption{Continued.	}
	\label{fig:fits2}
\end{figure*}   


\subsubsection{Inputs for pPXF Velocity Dispersion Calculations}

When available, spectroscopy from both the $H$- and $K$-bands was used. 
Observed spectra were logarithmically rebinned and corrected for instrumental resolution, and templates were resampled to match this resolution.
In cases where the resolution for a template was less than that of the observed spectra, the spectra were binned using inverse variance weighting to match the model resolution.
Numerous runs were performed for each galaxy using a variety of spectral template libraries, wavelength masking strategies, and a range of additive Legendre polynomial orders to limit the effects of template mismatch and telluric correction inaccuracies.
The variety of inputs also allows us to characterize the systematic error on the velocity dispersion, which exceeds the statistical error provided by pPXF.

Extensive testing of pPXF on a sample of five massive quiescent galaxies $1.4<z<2.1$ was performed in \citet{vandeSande2013}. In their Appendix A, they test the dependence of velocity dispersions output by pPXF on various inputs, including template choice, polynomial degree, and stellar population models.
We do similar testing when fitting the \magazine\ sample, which is described in more detail in Appendix~\ref{App:pPXF}.

Briefly, we ran pPXF using the templates from \citet[][BC03]{Bruzual2003}, SSPs constructed from the MILES library \citep{Sanchez-Blazquez2006, Vazdekis2010}, and the Indo-US library \citep{Valdes2004}.
These libraries provide sufficient variety of spectral templates to fit the observed spectra well.
However, pPXF does not incorporate galaxy photometry into the fit, and failure to do this can result in underestimating the velocity dispersions \citep[][though results are often consistent within the errors]{Mendel2020}.
As such, we also use FAST++ \citep{Schreiber2018a} with the \citet[][BC03]{Bruzual2003} templates to jointly fit the observed photometry and spectroscopy and obtain a best-fit template, subsequently using pPXF with that template choice fixed - we designate these runs as BC03++.

Runs with each of these four template sets (BC03++, BC03, MILES, and Indo-US) were also done with an additive Legendre polynomial from order $0\leq d \leq 50$.
Such a polynomial corrects for differences in template and observed spectral shape as can result from \eg\ telluric correction inaccuracies and helps avoid template mismatch.
The effect of adding a polynomial of very high order is to perturb a template to fit all the noise features in an observed spectrum, and thus a somewhat low order polynomial is preferred.
Choice of polynomial order varies in the literature: \citet{vandeSande2013} use $d\sim17$, with velocity dispersions only showing a small dependence on this choice from $0 \leq d \leq 50$, \citet{Mendel2020} use $d=9$,  \citet{Saracco2020} use $d=4$, \citet{Tanaka2019} use $d=1$, and \citet{Esdaile2021} do not use an additive polynomial (effectively $d=0$).
In general we find that the velocity dispersion varies the least over the range $10<d<20$ for the UMGs in this sample.

Finally, we also choose various methods of masking the spectral wavelengths used in the fit.
We test pPXF while masking all observed emission lines as well as: 1) all Balmer features, 2) the \Hbeta\ feature, 3) no other wavelengths, 4-6) wavelengths in 1-3 plus sky lines.
Exclusion of the Balmer features can result in a more stable velocity dispersion \citep{vandeSande2013} and remove any degeneracy between small scale emission and template choice, but also remove a strong constraint on the velocity dispersion for spectra with low SNR as is typical for galaxies at these redshifts \citep{Tanaka2019, Esdaile2021}.
Masking only the \Hbeta\ feature in these quiescent galaxies strongly mitigates the emission issue.

\subsubsection{Measured Velocity Dispersions}

Resultant best-fit templates from each run were visually inspected and also compared to the galaxy photometry, with results involving clearly incorrect templates discarded (these were uncommon, on the order of a few percent).
Our galaxies have sufficient SNR such that the results of the many runs form a distribution with a clear mode for each galaxy, which we use as the velocity dispersion in subsequent analysis.
The (asymmetrical) spread of the distribution of results is used to derive errors on the velocity dispersion, which can differ from the output error of pPXF by up to a factor of $\sim2$.
Median values of the fitted velocity dispersion distributions and averages weighted by reduced $\chi^2$ and reported error are all statistically consistent with the mode of the distribution.
Models with the best-fit velocity dispersions are shown in Figure~\ref{fig:fits1}.
Plots showing the dependencies on choice of input parameters, as well as a more complete discussion are included in Appendix \ref{App:pPXF}.

\subsubsection{Aperture Correction of the Measured Velocity Dispersions}

For comparison with other measurements in the literature, we correct the measured velocity dispersions to velocity dispersions at the effective radius, $\sigma_{\rm e}$ (size calculations are described in Section~\ref{Sec:SizeCalc}).
This removes instrumental dependence and accounts for the effects of seeing.
Such a correction is dependent upon the size and shape of the spectral aperture, the observing conditions (\ie\ seeing) and the size of the target.
The MOSFIRE aperture size of interest, $r_\textrm{aperture}$, is the distance along the slit over which the 1D spectrum was optimally extracted, and is thus a function of both intrinsic size and seeing conditions which varies for different masks on which the same object is located.
In theory, this could also be affected by the length of a slit if it was insufficiently long to cover the entire object (minimum MOSFIRE slit length is 7.1"), though this would only be a concern for very large objects or extremely poor conditions, which does not affect this sample.

Extensive modeling in Appendix B of \citet{vandeSande2013} shows that for a rectangular aperture with weighted extraction, this correction factor is quite flat as a function of $r_\textrm{aperture}/r_\textrm{eff}$, when the PSF is taken into account.
Indeed, the correction factors for our velocity dispersions calculated following \citet{vandeSande2013} range from 1.048 to 1.058, though the small differences in this correction are far exceeded by the errors on the measured velocity dispersions.
The corrected values are shown in Table \ref{tab:props} and used for the remainder of this analysis.

\subsection{Sizes}\label{Sec:SizeCalc}

GALFIT \citep{Peng2002, Peng2010} was used to model the $K_{\rm S}$-band images of all objects, and the $HST/WFC3/$F160W images of COS-DR3-201999, COS-DR3-202019, and COS-DR3-84674.
For the sources in the COSMOS field, the UltraVISTA DR4 $K_{\rm S}$ mosaic with pixel scale 0.15$^{\prime \prime}$ per pixel and FWHM=0.78$^{\prime \prime}$ \citep{McCracken2012} was adopted.
For the sources in the XMM field, the VIDEO DR4 $K_{\rm S}$ mosaic with pixel scale 0.2$^{\prime \prime}$ per pixel and FWHM=0.82$^{\prime \prime}$ \citep{Jarvis2013} was adopted.
The fitting process was similar for all the galaxies.
A small cutout centered on the relevant galaxy was created, making sure to include the central object and any nearby objects along with enough empty region for the sky background calculation.
In most cases, the central galaxy was fitted simultaneously with the neighboring objects.
In a few cases, the neighboring objects were not fitted if they were far enough from the UMG that their light was not contaminating the objects.
In this case, the neighboring objects were only masked out in the GALFIT fitting.
All objects were fitted with a single S\'ersic profile.
The free parameters had the following fitting constraints: the centroid of the object was allowed to vary at most by 2 pixels in each direction from the initial coordinates; $0.05 \leqslant r_{\rm e} [^{\prime \prime}] \leqslant 1$; $0.2 \leqslant n \leqslant 7$; and $0.1 \leqslant q \leqslant 1$.
We allowed GALFIT to fit a constant sky background as a free parameter.
Previous studies have shown this to be the preferred choice, and that GALFIT performs significantly better when allowed to internally measure a sky background, as opposed to being provided a fixed background \citep{Haussler2007, Cutler2022}.
Furthermore, the convolution box was allowed to span the whole cutout.

\begin{figure*}[h]
	\centering{\includegraphics[width=\textwidth,trim=0in 0in 0in 0in, clip=true]{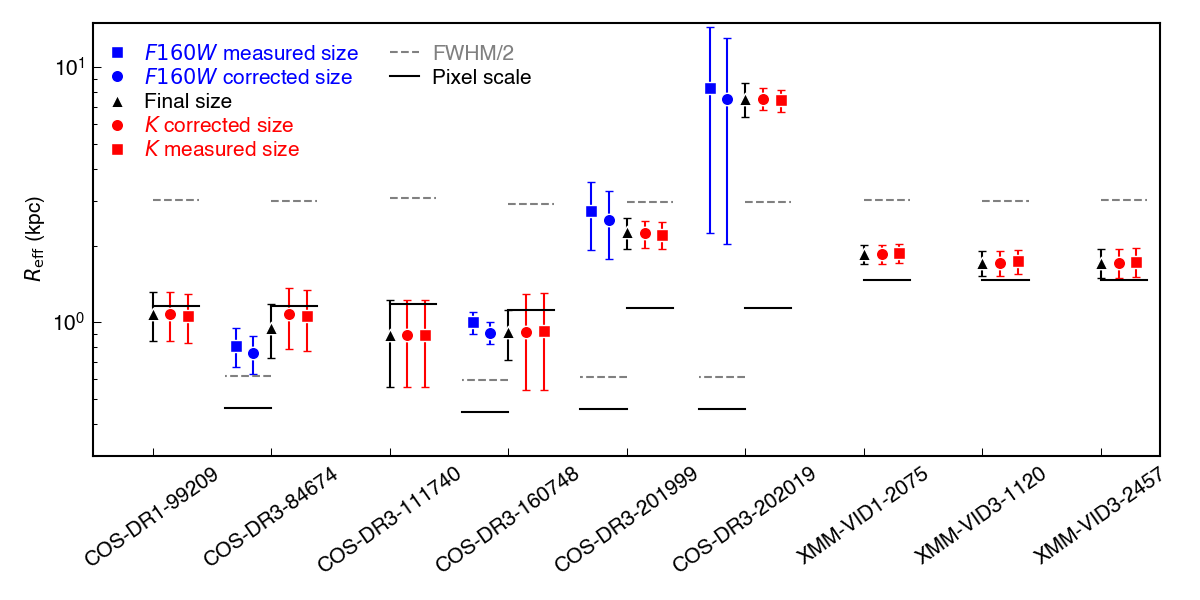}}
	\caption{Effective radii for the UMGs. Radii measured directly from the $K$-band and $F160W$ imaging are shown as squares colored red and blue, respectively. Correcting these sizes to 5000${\rm	\AA}$ radii results in the circles of the same colors. The weighted average of the corrected measurements (the same as the $K$-band value if $F160W$ data unavailable) is shown as a black triangle. The pixel size of the imager at the redshift of each target is shown as a black line, while the FWHM/2 is a gray dashed line.}
	\label{fig:sizecorr}
\end{figure*}

For each $K_{\rm S}$-band object fit, two to three nearby, unsaturated, uncontaminated, and background-subtracted stars were used as point-spread functions (PSFs) for model convolution.
We also adopted as the model PSF a high signal-to-noise PSF constructed using 10 different nearby stars, stacking the corresponding sky-subtracted stamps after masking any nearby objects, re-centering the stars, and normalizing the integrated flux.
Utilizing different stars/PSFs allows for a more realistic estimate of the size measurement error, which is generally underestimated by GALFIT.
For the $WFC3$/F160W images, a position-dependent PSF model was created using \texttt{grizli}  (https://grizli.readthedocs.io) to shift and drizzle HST empirical PSFs \citep{Anderson2015} at the position of the UMGs.
\citet{Cutler2022} showed that there is no significant difference in GALFIT structural measurements between galaxies fit with position-dependent PSFs and those with PSFs determined over a larger area of the mosaic, even at $z>2$.
While most of the galaxies are not resolved in the ground-based imaging, GALFIT can still recover fits down to FWHM/2 \citep{Haussler2013, Nedkova2021}, although \citet{Ribeiro2016} suggests such measurements tend to be underestimates.
Additionally, the sizes derived from the unresolved $K_{\rm S}$-band and the resolved $WFC3$/F160W GALFIT modelings are consistent with each other, as shown in Figure~\ref{fig:sizecorr}.

The SNR of the images is not sufficiently high to obtain a reliable value of the S\'ersic index.
As the size and S\'ersic index are covariant in the fitting process, we also use GALFIT to perform fits with the S\'ersic index fixed to $n=1,3,4,6$ and compare to the reported best fit, in which n is allowed to vary, to discern another source of possible error on the size measurement.
In some cases, these fits do not converge, and in some the reported fit is clearly incorrect upon visual inspection.
Ignoring these cases, we find that objects with best-fit S\'ersic index $2<n<4$ from ground-based $K$-band imaging show size variations on the order of 10\% in these tests.
In the other cases, variation on the order of up to 20\% is seen.
For all UMGs, these variations are smaller than the reported errors based on different characterizations of the PSF.


For galaxies with imaging in both bandpasses, the measured sizes are consistent within the errors.
However, the two bandpasses are probing different wavelengths, which can be on opposite sides of the \Dfour\ feature.
To avoid any issues on this front, we convert all measured sizes to rest-frame 5000$\rm \AA$ sizes
following \citet{vanderWel2014} as:
\begin{eqnarray}
r_{\rm eff, 5000\AA} = r_{\rm eff, \lambda_{\rm obs}} \bigg{(}\frac{1+z}{1+z_{\rm pivot}}\bigg{)}^\frac{\Delta {\rm log} (r_{\rm eff})}{\Delta {\rm log} \lambda}
\end{eqnarray}
where,
\begin{eqnarray}
z_{\rm pivot} &=& \lambda_{\rm obs} / 5000{\rm \AA} - 1\\
\frac{\Delta {\rm log} (r_{\rm eff})}{\Delta {\rm log} \lambda} &=& -0.35 + 0.12z -0.25{\rm log}\bigg{(}\frac{M_*}{10^{10}M_\odot}\bigg{)}
\end{eqnarray}

For rest-frame optical sizes (5000${\rm \AA}$), $z_{\rm pivot}(F160W) = 2.2$ and $z_{\rm pivot}(K) = 3.3$.
While the $\frac{\Delta {\rm log} (r_{\rm eff})}{\Delta {\rm log} \lambda}$ relation from \citet{vanderWel2014} was derived using less massive galaxies at $z<2$, we note that the corrections here are considerably smaller than the errors on the size measurements.
Given the consistency of all these half-light radii for a given galaxy, in what follows we use a weighted average of the corrected size measurements in all available bands for determination of dynamical mass.
This size is listed in Table \ref{tab:props}.

We also note that the morphology of COS-DR3-201999 was analyzed in \citet{Lustig2020}, with the id~252568, which returned 
$r_{\rm e}$(5000 ${\rm \AA}$)/kpc = $2.37^{+0.58}_{-0.37}$,
a size fully consistent with the analysis herein.

\subsection{Dynamical Masses}

The velocity dispersion and effective radius measurements can be used to calculate dynamical masses for the UMGs in this sample, 
\begin{eqnarray}
M_{\rm dyn}(<r_{\rm e})&=& \kappa_{\rm e}\frac{\sigma_{\rm e}^2 r_{\rm e}}{G},
\end{eqnarray}
where $\kappa_{\rm e}$ is a virial coefficient which depends upon the (an-)isotropy of the stellar velocities and the intrinsic mass profile of the galaxy.
This value has been calibrated using lower redshift ellipticals, as such determinations for high-redshift, compact quiescent galaxies have not been done due to their small sizes and faint magnitudes.
The typical value used for $z\sim2$ quiescent galaxies is $\kappa_{\rm e}=2.5$ \citep{Newman2012,Barro2014}.
The resultant value of $M_{\rm dyn}(<r_{\rm e})$ is then doubled to estimate the total $M_{\rm dyn}$, which is then compared to the total stellar mass.

\citet{Cappellari2006} also published an analytical estimator which folds in both the virial coefficient and the correction to total mass
\begin{eqnarray}
M_{\rm dyn}&=&\beta (n) \frac{\sigma_{\rm e}^2 r_{\rm e}}{G}\\
\beta (n) &=& 8.87-0.831n+0.0241n^2
\end{eqnarray}
For a sample of massive, quiescent galaxies at $z\sim2$, a typical value of $\beta (n) \sim5$ is found, which is equivalent to the choice of $\kappa_{\rm e}=2.5$ \citep{vandeSande2013, Belli2014a}.
Previous samples of UMGs at \mbox{$z\gtrsim3$} \citep{Esdaile2021, Saracco2020} have used this estimator and returned values in the range of $5.4<\beta (n)<6.4$, while \citet{Tanaka2019} also adopt $\beta (n)=5$ due to a lack of a confident measure of S\'ersic index.

In this work we also adopt the value of $\beta (n)=5$, as the SNR of the images used for size calculations is not sufficiently high to obtain a reliable value of the S\'ersic index.
Results of these calculations are provided in Table~\ref{tab:props}.


\begin{figure*}[tp]
	\centering{\includegraphics[width=\textwidth,trim=0in 0in 0in 0in, clip=true]{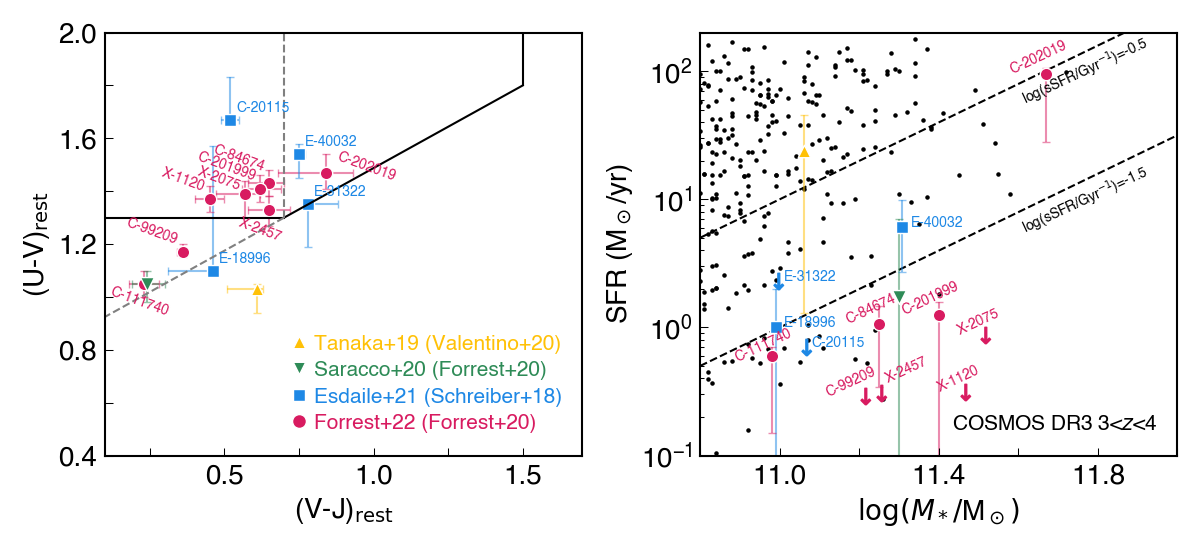}}
	\caption{Properties of the massive quiescent galaxies in the $z\gtrsim3$ sample which have published stellar velocity dispersions.
	 \textit{Left}: The rest-frame color $UVJ$ diagram. Galaxies are labeled here and in subsequent plots by the reference containing stellar velocity dispersion measurements. Rest-frame colors, star-formation rates, and/or stellar masses are often from separate publications, also noted. Data at $z\gtrsim3$ are shown as: a gold upward facing triangle \citep{Tanaka2019}, a seagreen downward facing triangle \citep{Saracco2020}, azure squares \citep{Esdaile2021}, and raspberry circles (this work).
	 \textit{Right:} The relation between star-formation rate and stellar mass. Galaxies with photometric redshifts $3<z<4$ from the COSMOS-UltraVISTA DR3 catalog are shown as black points, and dashed lines are plotted showing constant specific SFR (-1.5 and -0.5 Gyr$^{-1}$), corresponding roughly to the main sequence and one dex below it.}
	\label{fig:colors}
\end{figure*}


\begin{table*}
  \centering
  \caption{Properties of massive quiescent galaxies in the $z\gtrsim3$ sample discussed in this work.}
    \begin{tabular}{ llcclcc }
      UMG  & \zspec & \logM & \logMdyn & $\sigma_e$ (km s$^{-1}$) & $r_\textrm{eff, 5000{\rm \AA}}$ (kpc) & Reference \\
      \hline \hline
     COS-DR1-99209      & $2.9834^{+0.0023}_{-0.0028}$       & $11.22^{+0.05}_{-0.06}$    & $11.31^{+0.12}_{-0.20}$   & $401^{+63}_{-84}$ 	   & $1.08\pm0.24$		&  This work    \\
     COS-DR3-84674		& $3.0094^{+0.0015}_{-0.0011}$	& $11.25^{+0.01}_{-0.02}$     & $11.33^{+0.23}_{-0.14}$   & $442^{+206}_{-68}$	& $0.95\pm0.23$	&  This work    \\
     COS-DR3-111740	& $2.7988^{+0.0013}_{-0.0011}$	& $10.98^{+0.01}_{-0.00}$    & $11.02^{+0.13}_{-0.24}$   & $467^{+102}_{-131}$	& $0.89\pm0.33$ 	&  This work  \\
     COS-DR3-201999	& $3.1313^{+0.0014}_{-0.0012}$	& $11.40^{+0.03}_{-0.01}$     & $11.28^{+0.15}_{-0.24}$   & $271^{+55}_{-58}$		& $2.26\pm0.31$	&  This work   \\
     COS-DR3-202019	& $3.1326^{+0.0021}_{-0.0011}$	& $11.67^{+0.04}_{-0.05}$     & $12.00^{+0.14}_{-0.27}$    & $345^{+92}_{-111}$	& $7.54\pm1.16$	&  This work  \\
     XMM-VID1-2075		& $3.4520^{+0.0014}_{-0.0017}$	& $11.52^{+0.00}_{-0.05}$     & $11.49^{+0.12}_{-0.11}$    & $379^{+85}_{-53}$		& $1.85\pm0.16$	&  This work   \\
     XMM-VID3-1120		& $3.4919^{+0.0018}_{-0.0029}$	& $11.47^{+0.02}_{-0.03}$     & $11.54^{+0.10}_{-0.31}$    & $419^{+74}_{-148}$	& $1.71\pm0.19$	&  This work  \\
     XMM-VID3-2457		& $3.4892^{+0.0032}_{-0.0024}$	& $11.26^{+0.02}_{-0.03}$     & $11.49^{+0.07}_{-0.29}$    & $396^{+40}_{-132}$	& $1.71\pm0.22$	&  This work  \\
     ZF-COS-20115		& $3.715$       					& $11.06^{+0.06}_{-0.04}$     & $10.86^{+0.14}_{-0.20}$	& $283^{+52}_{-52}$		& $0.66\pm0.08$	& \citet{Esdaile2021} \\
     3D-EGS-40032		& $3.219$       					& $11.31^{+0.03}_{-0.03}$     & $11.41^{+0.11}_{-0.16}$	& $275^{+56}_{-56}$		& $2.40\pm0.19$	& \citet{Esdaile2021} \\
     3D-EGS-18996		& $3.239$       					& $10.99^{+0.02}_{-0.03}$     & $10.56^{+0.13}_{-0.19}$	& $196^{+48}_{-48}$		& $0.63\pm0.05$	& \citet{Esdaile2021} \\
     3D-EGS-31322		& $3.434$       					& $10.99^{+0.05}_{-0.04}$     & $10.85^{+0.20}_{-0.39}$	& $201^{+119}_{-119}$	& $0.61\pm0.05$	& \citet{Esdaile2021} \\
     C1-23152$^a$    		& $3.352^{+0.002}_{-0.002}$		& $11.30^{+0.19}_{-0.13}$     & $11.34^{+0.07}_{-0.09}$   & $409^{+60}_{-60}$	   	& $1\pm0.1$ 		& \citet{Saracco2020} \\
     SXDS-27434	   	& $4.0127^{+0.0005}_{-0.0005}$	& $11.06^{+0.04}_{-0.04}$     & $<11.32$				& $268^{+59}_{-59}$	  	& $<1.3$			& \citet{Tanaka2019} \\
      \hline
    \end{tabular}\\
    \footnotesize{$a$. This object is renamed COS-DR3-160748 in the MAGAZ3NE Survey.}
    \label{tab:props}
\end{table*}

\section{Results \& Discussion} \label{Sec:Res}

We compare our results to massive, quiescent galaxies at a range of redshifts.
The first sample, from \citet{Posacki2015}, is a reanalysis of 55 massive early-type galaxies at $z\sim0.2$ from SLACS \citep{Treu2010} and a subset of 223 \Hbeta\ massive absorption line galaxies in the local volume from ATLAS$^{\rm 3D}$ \citep{Cappellari2013b}.
Galaxies selected from SDSS with velocity dispersions $\sigma>350$~km/s at similar redshifts were also compared in an attempt to mitigate progenitor bias \citep{Bernardi2006, Saracco2020}.
\citet{Mendel2020} compiled and reanalyzed spectra from early-type galaxies at $1.4<z<2.1$, including spectra presented in \citet{Cappellari2009, Newman2010, Toft2012, Bezanson2013a, vandeSande2013, Belli2014a, Belli2014b, Barro2016, Belli2017}.
In addition to our eight $z\gtrsim3$ UMGs, we fold in six previously published $z>3$ UMGs with velocity dispersion measurements: SXDS-27434 \citep{Tanaka2019}; C1-23152 (published by members of our group in \citet{Saracco2020} and subsequently renamed as COS-DR3-160748 in the context of the MAGAZ3NE Survey; \citet{Forrest2020b}); ZF-COS-20115, 3D-EGS-40032, 3D-EGS-18996, and 3D-EGS-31322 \citep{Esdaile2021}.

Galaxies in these works have also been studied spectroscopically in \citet{Valentino2020}; \citet{Marsan2015} and \citet{Forrest2020b}; \citet{Glazebrook2017} and \citet{Schreiber2018b}, respectively.
For the most part, the massive galaxies at $z>1.4$ were selected for spectroscopic follow-up via a combination of magnitude/stellar mass, color/SFR, and photometric redshift cuts.
Nonetheless it is important to keep in mind that these cuts are not identical given the different survey depths, photometric wavelength coverage, and photometric redshift tools.
Thus it is possible that studies are selecting different sub populations of massive quiescent galaxies.

In Figure~\ref{fig:colors}, we show the rest-frame colors, stellar masses, and star-formation rates of the objects in the $z\gtrsim3$ sample.
Most of the galaxies are consistent with recently quenched post-starburst galaxies, as they lie in the lower left of the $UVJ$ quiescent wedge or slightly blueward of it and show SFRs significantly below the main sequence for their mass at this redshift.
The two notable exceptions to this are COS-DR3-202019 and SXDS-27434 \citep{Tanaka2019}.
The former has \mbox{$SFR=82$~$M_\odot$/yr} and is the reddest and most massive of the sample, consistent with a dusty star-forming galaxy \citep{Forrest2020b}, while the latter has \mbox{$SFR=24$~$M_\odot$/yr} and has the bluest $(U-V)_{\rm REST}$ color of the sample \citep{Valentino2020}.


\begin{figure}[tp]
	\centering{\includegraphics[width=0.5\textwidth,trim=0in 0in 0in 0in, clip=true]{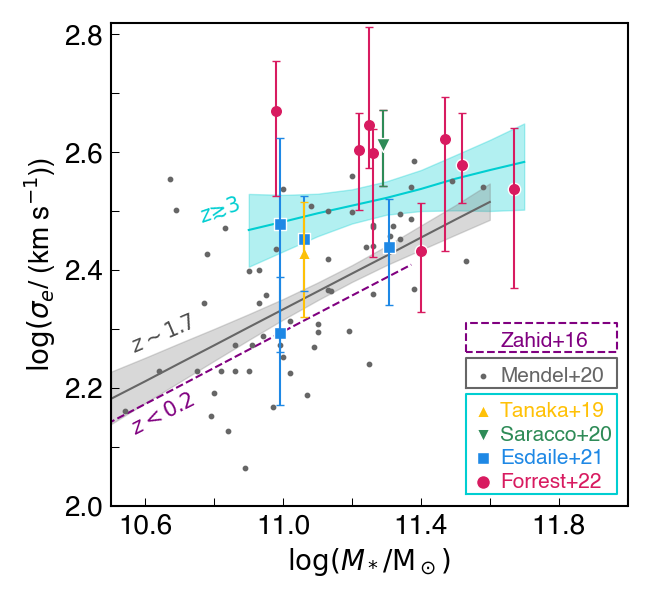}}
	\caption{Stellar velocity dispersions and stellar masses of the massive quiescent galaxies. The colors and markers used for the $z\gtrsim3$ sample are identical to those in Figure~\ref{fig:colors}, while data from \citet{Mendel2020} at $1.4<z<2.1$ are shown in gray. The median error bar for the \citet{Mendel2020} dataset is $\sim0.1$ dex. Fits to both sets of data are shown as turquoise and gray lines, respectively, with shaded error regions representing the $16^{th}$ to $84^{th}$ percentile range from Monte Carlo resampling. The fit to low redshift data from \citet{Zahid2016} is shown in purple.}
	\label{fig:VDmass}
\end{figure}


\subsection{Large Velocity Dispersions}

The best-fit velocity dispersions for the MAGAZ3NE sample are very large, at $\sim400$ km s$^{-1}$.
Nonetheless, several galaxies at $z\sim2$ have previously been measured with similarly high velocity dispersions \citep{vanDokkum2009a, vandeSande2013, Belli2014b, Belli2017}.
These velocity dispersions confirm the large stellar masses of these objects while being independent of the various problems intrinsic to SED fitting such as choice of IMF and contamination by emission lines (see Section~\ref{Sec:MassProbs}).

A positive correlation between stellar velocity dispersion and stellar mass is expected as the mass within the small sizes over which we probe stellar velocity dispersion is dominated by stars.
At $1.4<z<2.1$, the data compiled in \citet{Mendel2020} show a positive correlation between the two, though individual galaxies show significant scatter. 
A least-squares regression to the entire $z\gtrsim3$ sample shows a vertical offset towards larger velocity dispersions at a given stellar mass, but a similar slope to both the $1.4<z<2.1$ sample and a sample of  massive quiescent galaxies from SDSS \citep{Zahid2016}, shown in Figure~\ref{fig:VDmass}.


\begin{figure}[tp]
	\centering{\includegraphics[width=0.5\textwidth,trim=0in 0in 0in 0in, clip=true]{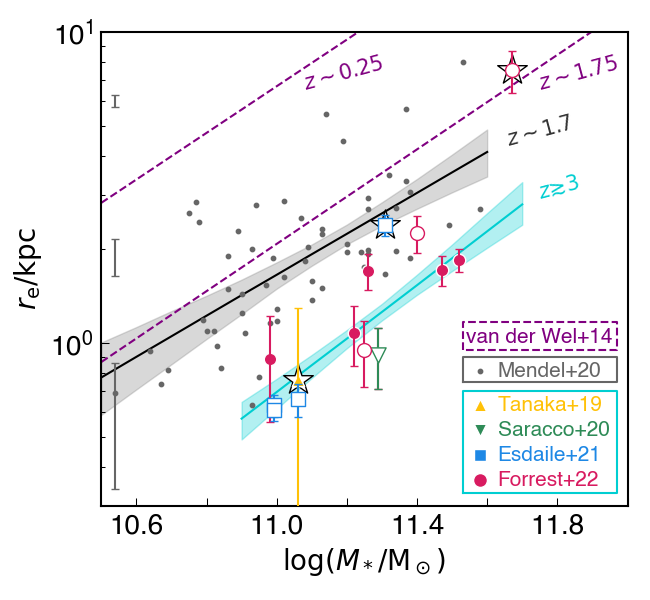}}
	\caption{Effective radii and stellar masses of the massive galaxies. Colors and markers remain the same as in Figure~\ref{fig:VDmass}, and objects with sizes derived from high-resolution $HST$ imaging have white centers.
	Representative error bars for the \citet{Mendel2020} dataset are shown on the left.
	The three galaxies with $SFR>3$~M$_\odot$/yr are marked with a black star and excluded from the $z\gtrsim3$ fit.
	The size-mass relations for early-type galaxies at $z\sim0.25$ and $z\sim1.75$ from \citet{vanderWel2014} are shown as purple dashed lines, while fits to the \citet{Mendel2020} dataset and the $z\gtrsim3$ dataset are a black line and a solid turquoise line with the range of Monte Carlo fits from $16-84\%$ shaded accordingly. }
	\label{fig:sizemass}
\end{figure}


\subsection{The Size-Stellar Mass Relation}

At a given epoch, the effective radius and stellar mass of a galaxy are also correlated, though quiescent and star-forming galaxies tend to follow different relations, and those relations evolve with time to smaller sizes for a given stellar mass at earlier times \citep[\eg][]{vanderWel2014, Straatman2015a, Mowla2019a, Marsan2019}.

Indeed, the $1.4<z<2.1$ sample from \citet{Mendel2020} is in agreement with the $z\sim1.75$ relation for early-type galaxies presented in \citet{vanderWel2014} using data from 3D-HST:
\begin{eqnarray*}
r_{\rm e}/{\rm kpc}
&=& 1.23 \times (M_*/5\times10^{10}M_\odot)^{0.76},
\end{eqnarray*}
or equivalently,
\begin{eqnarray*}
\log(r_{\rm e}/{\rm kpc}) &=& -8.04+0.76\log(M_*/M_\odot)
\end{eqnarray*}
From Monte Carlo resampling of the $z\gtrsim3$ galaxies with $SFR<3$~M$_\odot$/yr we find
\begin{eqnarray*}
\log(r_{\rm e}/{\rm kpc}) &=& -9.73(\pm 1.50)+0.87 (\pm 0.15) \log(M_*/M_\odot),
\end{eqnarray*}
that is, smaller sizes for a given stellar mass showing a statistically consistent, but perhaps slightly steeper relation with stellar mass (see Figure~\ref{fig:sizemass}).

Relative to the $z\sim1.75$ relation, this $z\gtrsim3$ fit shows smaller sizes by a factor greater than 3 at \logM$\sim11$ and a factor of $\gtrsim2$ at \logM$\sim11.5$, which also agrees with the redshift size evolution shown in \citep{Straatman2015a}.
Limiting the $z\gtrsim3$ sample to quiescent galaxies with $HST/WFC3$ imaging does not significantly change the best-fit relation, though including the galaxies with $SFR>3$~M$_\odot$/yr does result in a steeper slope.

\subsection{Comparison of Dynamical Mass and Stellar Mass}

\begin{figure}[tp]
	\centering{\includegraphics[width=0.5\textwidth,trim=0in 0in 0in 0in, clip=true]{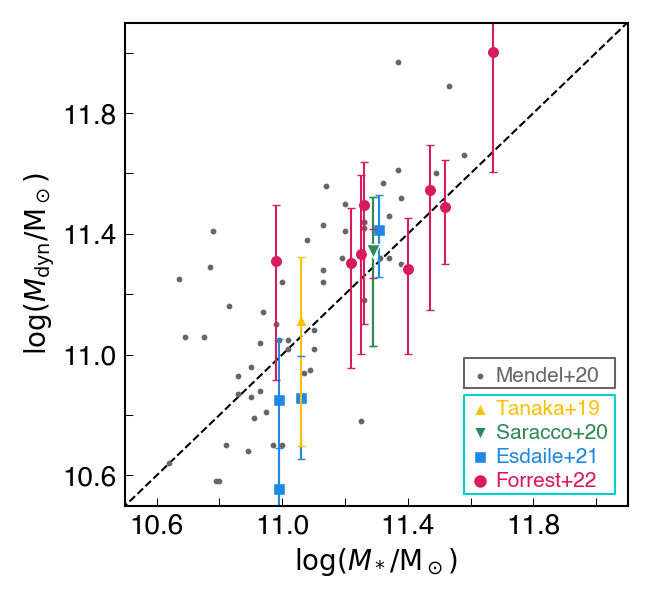}}
	\caption{Dynamical and stellar masses of the massive quiescent galaxies. The colors and markers used are identical to those in Figure~\ref{fig:sizemass}. The black dashed line represents a ratio of unity, corresponding to a Chabrier IMF. }
	\label{fig:sdmass}
\end{figure}

The dynamical and stellar masses for the $z\gtrsim3$ sample are listed in Table~\ref{tab:props} and shown in Figure~\ref{fig:sdmass}.
For massive, quiescent galaxies with little gas or dust and small sizes, the dynamical and stellar masses are expected to be quite similar as the central regions are dominated by baryons with little dark matter contribution.
The most obvious exception to this in the MAGAZ3NE sample is COS-DR3-202019 (the most massive galaxy in the sample), which has a  radius $\sim3\times$ larger than any other galaxy in the sample, and is also the only one that shows evidence of ongoing star formation (see Figure~\ref{fig:colors}), but is still consistent with a 1-to-1 ratio between stellar and dynamical mass within $1\sigma$.
The consistency of the $z\gtrsim3$ sample's ratios of dynamical to stellar mass, \MDS, with unity suggests that the Chabrier IMF used to derive the stellar masses for these objects is in general reasonable.

While similarly massive galaxies at lower redshifts appear to prefer heavier IMFs \citep[\eg][]{Conroy2012, Cappellari2013a, Zahid2017}, at $z\sim1.7$ \citet{Mendel2020} also find that a lighter IMF such as the Chabrier IMF is required to prevent stellar masses from exceeding dynamical masses.
Dynamical masses in significant excess of stellar masses would be expected if either the choice of IMF is incorrect or if there is an appreciable fraction of dark matter in the galaxy.
We note that the contribution of dark matter for similar galaxies at lower redshift, $\sim 5-20\%$ \citep{Cappellari2013b, Mendel2020}, is too small to be quantified here given the observational errors involved.

\begin{figure*}[tp]
	\centering{\includegraphics[width=\textwidth,trim=0in 0in 0in 0in, clip=true]{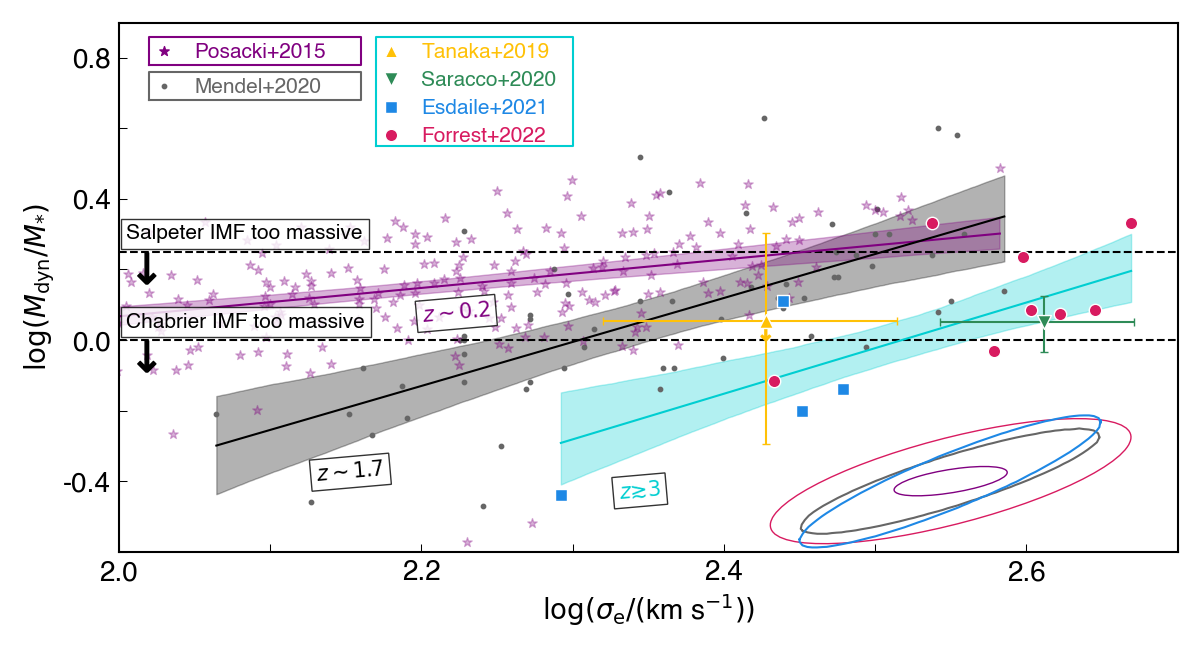}}
	\caption{Ratio of dynamical-to-stellar mass (IMF mismatch parameter) plotted against velocity dispersion at the half light radius.  The best fit to the data from \citet{Posacki2015} at $z\sim0.2$ is shown in purple, while the best fit to the data from \citet{Mendel2020} at $1.4<z<2.1$ is shown in gray. Data at $z>3$ are shown in the same colors as the previous plot, while the fit is in turquoise. Average error ellipses for the datasets from \citet{Posacki2015}, \citet{Mendel2020}, \citet{Esdaile2021}, and this work are shown in the bottom right. Horizontal dashed lines show the result when using a Salpeter or Chabrier IMF, below which using the respective IMF results in a stellar mass greater than the dynamical mass.}
	\label{fig:alpha}
\end{figure*}

That said, a comparison of \MDS\ to stellar velocity dispersion can still yield important insights.
For instance, high redshift quiescent galaxies have lower ratios of \MDS\ for a given velocity dispersion than galaxies at lower redshifts \citep{vandeSande2013, Hill2016, Belli2017, Mendel2020, Esdaile2021}, which is suggestive of a preference for a lighter IMF in such systems in early times.
While our data do not allow for significant constraints on dark matter content or IMF form for individual galaxies, a combination of the eight new MAGAZ3NE galaxies presented here with the four UMGs from \citet{Esdaile2021}, one from \citet{Saracco2020}, and one from \citet{Tanaka2019} allow the first look at these properties using a statistical sample at $z\gtrsim3$, shown in Figure~\ref{fig:alpha}.

We perform a linear regression between the logarithm of \MDS\ and the logarithm of the velocity dispersion at the effective radius for our sample, as well as those at $z\sim0.2$ and $z\sim1.7$.
Additionally, we use Monte Carlo resampling (accounting for the correlated errors) to characterize the uncertainties on the resulting best-fits:
\begin{eqnarray}
\log(M_{\rm dyn}/M_{*})_{z\sim0.2} & = & (0.29 \pm 0.02) +\\\nonumber
  && (0.40 \pm 0.05) \times \log(\sigma_{\rm e}/350)\\
\log(M_{\rm dyn}/M_{*})_{z\sim1.7} & = & (0.30 \pm 0.06) + \\\nonumber
  && (1.25 \pm 0.20) \times \log(\sigma_{\rm e}/350) \\
\log(M_{\rm dyn}/M_{*})_{z\gtrsim3} & = & (0.03 \pm 0.04) + \\\nonumber
  &&(1.29 \pm 0.36) \times \log(\sigma_{\rm e}/350) 
\end{eqnarray}

The best-fit slope at $z\gtrsim3$ ($1.29 \pm 0.36$) is consistent with that of the fit at $z\sim1.7$ ($1.25 \pm 0.20$) and significantly steeper than the low-redshift relation ($0.40 \pm 0.05$).
Additionally, the $z\gtrsim3$ sample is offset to lower \MDS\ by $\sim0.3$ dex relative to the $z\sim 1.7$ sample and $\sim0.5$ dex relative to the low redshift sample for a given velocity dispersion.
This means that while the $z\gtrsim3$ sample shows the same trend of preferring a heavier IMF at higher velocity dispersions relative to lower velocity dispersions, many of the highest velocity dispersion objects prefer a Chabrier IMF (or an IMF lighter than Chabrier) to a Salpeter IMF (see Figure~\ref{fig:alpha}).
In order for high velocity dispersion galaxies to prefer a bottom-heavy IMF such as Salpeter or even heavier \citep[\eg][]{Conroy2012}, at least one of several parameters must be systematically incorrect and provide a 0.2 dex ($\sim60\%$) gain in \MDS, addressed below.


\subsubsection{Are the high velocity dispersions too low?}

The reported velocity dispersions herein are some of the largest measured \citep[see also][for other galaxies with $\sigma>400$ km s$^{-1}$]{vanDokkum2009a, vandeSande2013, Saracco2020}.
To reach agreement with a Salpeter IMF, the velocity dispersions would have to be even higher by $\sim100$ km s$^{-1}$ for the highest velocity dispersion objects (and $\sim 500$ km s$^{-1}$ for those galaxies with lower velocity dispersions).
This increase is perhaps not unrealistic for some galaxies here given the errors on the measured velocity dispersions.
Intriguingly, this is in line with the large velocity dispersion of $510~\rm{km\ s}^{-1}$ measured for a massive, compact galaxy at $z=2.2$ in \citet{vanDokkum2009}.
Of course, while we have performed a robust investigation into the possible systematics involved in the calculation of velocity dispersions for this sample (see \mbox{Appendix~\ref{App:pPXF}}), the fact remains that the systematics may contribute to the results.

Another complicating factor here is the possibility of significant rotation in these systems, which would make the use of the measured velocity dispersion in the calculation of dynamical mass incorrect.
Several massive, quiescent galaxies at $z\sim2$ are disk-dominated and have been confirmed to have significant rotation thanks to gravitational lensing \citep{Toft2017, Newman2018b}.
Resolving rotation is not possible with our data.
Measured velocity dispersions could be inflated by a rotational component if a spectral slit is oriented with the major axis of the disk or could be underestimated if the spectral slit is misaligned.
Our sample is not large enough to claim that these effects cancel each other out on average.

\subsubsection{Are the size measurements too small?}

The GALFIT package used is widely used and appears to be generally accurate in calculating sizes.
Several of the objects in the $z\gtrsim3$ sample are not resolved in ground-based imaging, which can lead to incorrect size estimates below FWHM/2, particularly if the PSF is not well determined \citep{Haussler2013, Nedkova2021}.
Fortunately, the agreement between sizes calculated from $HST$ and ground-based imaging indicates that the sizes are reliable.
To find agreement with the $z\sim1.75$ relations from \citet{vanderWel2014} and the \citet{Mendel2020} dataset, the sizes must be $2-4\times$ larger than measured.
To improve consistency with a Salpeter IMF, the sizes must be underestimated by $\sim30\%$, which is considered unlikely as objects which are barely resolved are more likely to have their sizes overestimated.

\subsubsection{Are the dynamical masses calculated appropriately?}

The calculation of dynamical mass, in addition to relying upon accurate size and stellar velocity dispersion measurements, also contains a factor to account for the distribution of mass in the system.
The standard transformation used is a function of S\'ersic index, $n$.
While the imaging used to calculate sizes is not deep enough to reliably recover a S\'ersic index \citep[this usually requires SNR $\sim3\times$ deeper than that required for size measurements;][]{Haussler2007, Haussler2013}, the assumption of $\beta(n)=5$ corresponds to $n\sim5.5$.
An increase of $\sim60\%$ in this factor would require $n\sim1.2$, typically seen in larger galaxies with well developed disks, which remains a possibility.

The correction factor $\beta(n)$ was originally derived using low redshift elliptical galaxies, and while it shows great precision across $2<z<10$ \citep{Bertin2002, Cappellari2006}, it is possible such a transformation is not accurate for these galaxies for some reason.
Deeper, higher resolution imaging, as may be obtained with \textit{JWST} may allow for insights into this possibility.

\subsubsection{Are the stellar masses accurate?}\label{Sec:MassProbs}

Mass inaccuracies can be caused by the presence of strong emission lines, which can cause an overestimate of up to $\sim0.5$ dex \citep[\eg][]{Stark2013, Salmon2015, Forrest2017}.
However, \citet{Forrest2020b} model stellar masses for the MAGAZ3NE galaxies in this sample after correcting broadband photometry for any strong emission lines seen in the spectra (\OII, \OIII, \Hbeta), though of the sample here only COS-DR3-202019 shows significant emission.
The only other strong line which could normally be an issue is H$\alpha$, though at the redshifts of the sample, this line falls in between the $K$-band and the IRAC 3.6$\mu$m bandpass, and so should not affect the photometry either.

It is also known that the choice of modeling parameters and program can lead to differences in stellar mass calculations of around 0.2 dex \citep[\eg][]{Mobasher2015}.
\citet{Leja2019} compare stellar masses for objects in the 3D-HST study derived using FAST \citep{Kriek2009} and Prospector-$\alpha$ \citep{Leja2017}, and while they find a systematic offset of up to 0.4 dex in stellar mass, these differences appear to be $<0.1$ dex for high mass galaxies at high redshifts, as is the case for our sample.
Regardless, the Prospector-$\alpha$ code outputs higher stellar masses than FAST, and thus any such offset would only increase the tension with \eg\ a Salpeter IMF.

The possibility of young stars outshining older populations in a spectrum and leading to a light-weighted stellar mass different from the true stellar mass would similarly result in an \textit{underestimate} of the stellar mass.

Of course, the calculation of stellar masses rests upon the modeling of stellar populations, often based on the spectra of local stars.
It is possible that these model populations are not applicable to stellar populations in the early Universe.
A test of this possibility would require high-resolution stellar spectra at high redshifts and is not currently technologically feasible.


\begin{figure*}[tp]
	\centering{\includegraphics[width=\textwidth,trim=0in 0in 0in 0in, clip=true]{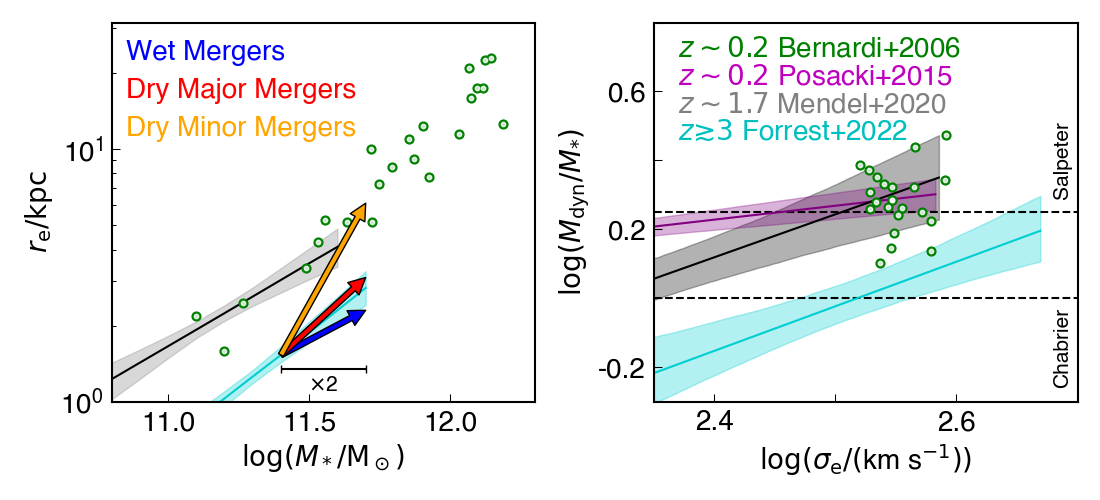}}
	\caption{Similar to Figures~\ref{fig:sizemass} and \ref{fig:alpha}, with additional low-redshift, high-velocity dispersion ETGs \citep[green circles;][]{Bernardi2006}.
	  The arrows on the size-stellar mass relation (left) show the effect on those properties via various merger scenarios, normalized to a doubling of the stellar mass.}
	\label{fig:bernardi}
\end{figure*}


\subsection{Evolutionary Insights}

The high velocity dispersions presented here for the $z\gtrsim3$ sample support the large stellar masses calculated through SED modeling for massive, high redshift galaxies and suggest that SED modeling of large photometric samples can be trusted to first order, outliers notwithstanding.
We also note however, that the velocity dispersions do not support much more mass than the stellar mass, implying that the contribution of dark matter at the centers of these compact galaxies is small.

\subsubsection{Progenitor Bias}

Galaxies with velocity dispersions such as those measured for some of our UMGs are exceedingly rare in the local Universe.
While analyzing the apparent trends seen in previous figures, we must carefully consider factors such as progenitor bias as well as the possibility that descendants of the rare $z\gtrsim3$ UMGs do not exist in the limited local volume.
In an attempt to mitigate these effects, we compare the UMGs in the $z\gtrsim3$ sample to an additional sample of massive low-redshift ETGs which are among the most massive galaxies in SDSS and which are not actively forming stars \citep{Bernardi2006, Saracco2020}. 
We correct the published velocity dispersions, originally corrected to $r_{\rm e}/8$, and transform them to $r_{\rm e}$ using the relation from \citet{Jorgensen1995},
\begin{eqnarray}
\frac{\sigma_{\rm ap}}{\sigma_{\rm e}/8} &=& \bigg{(}\frac{r_{\rm aper}}{r_{\rm e}/8}\bigg{)}^{-0.04}
\end{eqnarray}
which was used for the original correction in \citet{Bernardi2006} \citep[though see discussion about issues with this method for ETGs in][]{LaBarbera2019}.
This corresponds to a correction factor of $8^{-0.04}=0.92$.
Galaxies in this sample have larger stellar masses, velocity dispersions, and dynamical-to-stellar mass ratios than most of the $z\sim0.2$ sample from \citet{Posacki2015} (Figure~\ref{fig:bernardi}).
Velocity dispersion is known to correlate well with age for SDSS ETGs \citep[\eg][]{VanDerWel2009, Zahid2017}, and we thus assume that the stellar populations of these galaxies are also quite old.
We note that making cuts to the galaxy samples herein by stellar mass or velocity dispersion do not result in qualitative changes to our conclusions.

Spatially resolved studies of local massive ETGs have shown that stellar populations at their cores appear to be older than stars on the outskirts, as well as being regions with higher velocity dispersions \citep[\eg][]{vanDokkum2017, LaBarbera2019}, consistent with the bulk of star formation occurring at $z\gtrsim2$, followed by passive evolution via gas-poor (dry) mergers.
Dry major mergers, having a mass ratio between the two galaxies close to unity, increase both stellar mass and radius at similar rates with minimal new star formation (\ie\ retaining an old stellar age) and without much change in velocity dispersion \citep[\eg][]{Hopkins2009}.
Dry minor mergers on the other hand are expected to increase the effective radius approximately twice as fast as the stellar mass while also decreasing velocity dispersion slightly, though the cores of these galaxies could still retain high velocity dispersions \citep[\eg][]{Bezanson2009, Saracco2020}.

As seen in the left panel of Figure~\ref{fig:bernardi}, passive evolution of the $z\gtrsim3$ UMGs via dry minor mergers could lead to galaxies with sizes, stellar masses, and velocity dispersions of some of the most massive galaxies in SDSS \citep{Bernardi2006}.
While the $z\gtrsim3$ UMGs could evolve into galaxies at the massive end of the $z\sim1.7$ sample via dry minor mergers, those galaxies in the $z\sim1.7$ sample with lower stellar masses and velocity dispersions descend from galaxies with different properties than the $z\gtrsim3$ UMGs.
In particular, these progenitors have lower stellar masses and are possibly still forming stars at $z\sim3$.
Similarly, the $z\sim1.7$ sample could plausibly evolve into galaxies in the \citet{Bernardi2006} sample, but only those with larger velocity dispersions.

Most of the $z\gtrsim3$ UMGs herein are compact, post-starburst galaxies.
A gas-rich (wet) merger may have triggered such a burst of star formation \textit{in situ}, thus boosting the stellar mass significantly while keeping the effective radius small in contrast to the dry merger scenarios above \citep[\eg][]{Hopkins2009}.
Major mergers, wet or dry, are expected to be few in number for massive galaxies, and it is perhaps the case that the $z\gtrsim3$ UMGs have simply undergone additional major mergers relative to the progenitors of the lower mass half of the $z\sim1.7$ sample.
If so, the possibility exists that further major mergers would evolve these galaxies into systems more massive than any in the local volume.

Regardless, while the evolution of sizes, stellar masses, and velocity dispersions can be explained with mergers, the dynamical-to-stellar mass ratio is less easily explained.
If dynamical mass is calculated as $M_{\rm dyn} \propto r_{\rm e} \sigma_{e}^2$, then $\Delta \log(M_{\rm dyn}/M_{*})<0.05$ for all three merger cases described above.
Instead of an offset in dynamical-to-stellar mass at a given velocity dispersion, it may be that the increase in velocity dispersion from wet mergers is causing an offset in velocity dispersion at fixed dynamical-to-stellar mass, and we simply do not have any galaxies with large dynamical-to-stellar mass ratios in our $z\gtrsim3$ sample.
Major mergers can also introduce rotation into the system, making the dynamical mass calculations incorrect.

\subsubsection{IMF}

Detailed studies of absorption lines in local massive quiescent galaxies have suggested that the cores of these galaxies require a bottom-heavy ``super-Salpeter" IMF \citep[\eg][]{Lasker2013, Saulder2015, Conroy2017}. 
The higher inferred mass-to-light ratios associated with such a bottom-heavy IMF also correlate with velocity dispersion \citep[\eg][]{Conroy2013, Cappellari2013b}.
This is not only seen in samples of galaxies, but also within individual nearby galaxies, where the central cores have larger velocity dispersions and heavier inferred IMFs than the outskirts \citep[\eg][]{LaBarbera2019}.
While the sample of $z\gtrsim3$ galaxies in this study do show a similar trend towards a heavier IMF with higher velocity dispersion, none of the galaxies in our sample show evidence for a ``super-Salpeter" IMF and most require a Chabrier IMF in order for the stellar mass to remain below the dynamical mass, despite having very large velocity dispersions.
This creates a conundrum, as massive compact systems at high redshift, such as the $z\gtrsim3$ sample are generally thought to be the progenitors of the low redshift, high-mass sample such as that from \citet{Bernardi2006}, growing largely through mergers as described above \citep[\eg][]{Bezanson2009, VanDerWel2009, Saracco2020, Mendel2020}.
Such a picture does not offer a way to significantly change the observed IMF from high-redshift progenitors to the cores of local, massive ETGs.

An alternative possibility is that the IMF is determined by metallicity \citep[\eg][]{Koppen2007}, which shows a close positive correlation with inferred IMF slope for local ETGs from IFU data in the CALIFA survey \citep{Martin-Navarro2015}.
In this view, massive galaxies at early times undergo gas-rich mergers and form substantial fractions of their stars with gas containing a significant amount of metals from previous generations of stars.
This causes new star formation at high metallicity in the $z\gtrsim3$ UMGs \citep{Saracco2020} which occurs with a bottom-heavier IMF.
Meanwhile, less massive galaxies, being located in less massive halos, are more likely to build up their stellar mass not through merger-induced star formation, but by inflows of pristine gas.
Further, due to their lower masses, these galaxies lose many of the metals they produce via galactic outflows.
This then creates a lower-metallicity environment for star formation, which generates a bottom-lighter IMF.
This picture is also consistent with the mass-metallicity relationship seen for star forming galaxies out to $z>3$ \citep[\eg][]{Tremonti2004, Lian2018, Sanders2021}.
Subsequent growth of massive galaxies via minor mergers then deposits stars from the lower mass galaxies in the outskirts of the massive galaxy, producing the radial IMF trends seen in spatially resolved data \citep[\eg][]{vanDokkum2017, LaBarbera2019}.

\section{Conclusions}\label{Sec:Conc}

We have calculated stellar velocity dispersions and sizes for 8 UMGs at $z\gtrsim3$, more than doubling the sample at this epoch.
The high dispersions, on the order of $\sim400~\rm{km\ s}^{-1}$, are some of the largest measured, about $1.5\times$ those of galaxies at $z\sim1.7$ of similar stellar mass.
They also agree with the large stellar masses derived from SED fitting, supporting the conclusion that ultramassive quiescent galaxies at $z>3$ do exist in non-negligible numbers.
Size measurements for these objects additionally show a continuation of the evolution to smaller sizes at higher redshifts, with galaxies of similar stellar mass being about 1/3 the size of their $z\sim1.7$ counterparts.

We have used these size and stellar velocity dispersion measurements to calculate the dynamical mass.
The ratio of dynamical-to-stellar mass for these objects shows a trend with velocity dispersion as seen at lower redshifts, though it is offset to higher velocity dispersions / lower mass-to-light ratios.
This favors a Chabrier (or even bottom-lighter) IMF for most of the sample and is in tension with the ``super-Salpeter" IMFs seen in the cores of the most massive galaxies in the local Universe.

\section{Acknowledgements}
The authors wish to recognize and acknowledge the very significant cultural role and reverence that the summit of Maunakea has always had within the indigenous Hawaiian community.  We are most fortunate to have the opportunity to conduct observations from this mountain.
This work also uses data products from observations made with ESO Telescopes at the La Silla Paranal Observatory under ESO programme ID 179.A-2005 and on data products produced by CALET and the Cambridge Astronomy Survey Unit on behalf of the UltraVISTA consortium.

This material is based upon work supported by the National Science Foundation under Cooperative Agreement No. AST-2009442.
We gratefully acknowledge support from the NASA Astrophysics Data Analysis Program (ADAP) through grant numbers 80NSSC17K0019 and NNX16AN49G, the National Science Foundation through grants AST-1517863 and AST-2205189 and HST program numbers GO-15294 and GO-16300 provided by NASA through grants from the Space Telescope Science Institute, which is operated by the Association of Universities for Research in Astronomy, Incorporated, under NASA contract NAS5-26555.
Data presented herein were obtained using the UCI Remote Observing Facility, made possible by a generous gift from John and Ruth Ann Evans.
Some of the material presented in this paper is based upon work supported by the National Science Foundation under Grant No. 1908422.
P.S. acknowledges support by the grant PRIN-INAF-2019 1.05.01.85.11.

B.F. also thanks B. Lemaux, A. Pillepich, and L. Lewis for helpful discussions and input, as well as the authers of the codes referenced below, upon which this work has relied heavily.
Thanks to the anonymous referee as well, whose comments improved the manuscript.
 
\software{
Astropy \citep{Astropy2013,Astropy2018},
FAST++ \citep{Schreiber2018a},
\texttt{grizli} (ascl.net/1905.001),
IPython \citep{Perez2007},
Matplotlib \citep{Hunter2007},
Molecfit \citep{Smette2015, Kausch2015},
MosfireDRP (ascl.net/1908.007),
NumPy \citep{Oliphant2006},
pPXF \citep{Cappellari2017},
PyPeit \citep{Prochaska2020}
}

\bibliography{/Users/ben/Documents/library}

\begin{thebibliography}{}
\expandafter\ifx\csname natexlab\endcsname\relax\def\natexlab#1{#1}\fi
\providecommand{\url}[1]{\href{#1}{#1}}
\providecommand{\dodoi}[1]{doi:~\href{http://doi.org/#1}{\nolinkurl{#1}}}
\providecommand{\doeprint}[1]{\href{http://ascl.net/#1}{\nolinkurl{http://ascl.net/#1}}}
\providecommand{\doarXiv}[1]{\href{https://arxiv.org/abs/#1}{\nolinkurl{https://arxiv.org/abs/#1}}}

\bibitem[{Anderson {et~al.}(2015)Anderson, Bourque, Sahu, Sabbi, \&
  Viana}]{Anderson2015}
Anderson, J., Bourque, M., Sahu, K., Sabbi, E., \& Viana, A. 2015

\bibitem[{Ashby {et~al.}(2018)Ashby, Caputi, Cowley, Deshmukh, Dunlop,
  Milvang-Jensen, Fynbo, Muzzin, McCracken, F{\`{e}}vre, Huang, \&
  Zhang}]{Ashby2018}
Ashby, M. L.~N., Caputi, K.~I., Cowley, W., {et~al.} 2018, The Astrophysical
  Journal Supplement Series, 237, 39, \dodoi{10.3847/1538-4365/aad4fb}

\bibitem[{Barro {et~al.}(2014)Barro, Trump, Koo, Dekel, Kassin, Kocevski,
  Faber, {Van Der Wel}, Guo, P{\'{e}}rez-Gonz{\'{a}}lez, Toloba, Fang,
  Pacifici, Simons, Campbell, Ceverino, Finkelstein, Goodrich, Kassis,
  Koekemoer, Konidaris, Livermore, Lyke, Mobasher, Nayyeri, Peth, Primack,
  Rizzi, Somerville, Wirth, \& Zolotov}]{Barro2014}
Barro, G., Trump, J.~R., Koo, D.~C., {et~al.} 2014, Astrophysical Journal, 795,
  \dodoi{10.1088/0004-637X/795/2/145}

\bibitem[{Barro {et~al.}(2016)Barro, Kriek, P{\'{e}}rez-Gonz{\'{a}}lez, Trump,
  Koo, Faber, Dekel, Primack, Guo, Kocevski, Mu{\~{n}}oz-Mateos, Rujoparkarn,
  \& Seth}]{Barro2016}
Barro, G., Kriek, M., P{\'{e}}rez-Gonz{\'{a}}lez, P.~G., {et~al.} 2016, The
  Astrophysical Journal, 827, L32, \dodoi{10.3847/2041-8205/827/2/L32}

\bibitem[{Belli {et~al.}(2014{\natexlab{a}})Belli, Newman, \&
  Ellis}]{Belli2014a}
Belli, S., Newman, A.~B., \& Ellis, R.~S. 2014{\natexlab{a}}, Astrophysical
  Journal, 783, \dodoi{10.1088/0004-637X/783/2/117}

\bibitem[{Belli {et~al.}(2017)Belli, Newman, \& Ellis}]{Belli2017}
---. 2017, The Astrophysical Journal, 834, 18,
  \dodoi{10.3847/1538-4357/834/1/18}

\bibitem[{Belli {et~al.}(2014{\natexlab{b}})Belli, Newman, Ellis, \&
  Konidaris}]{Belli2014b}
Belli, S., Newman, A.~B., Ellis, R.~S., \& Konidaris, N.~P. 2014{\natexlab{b}},
  The Astrophysical Journal, 788, L29, \dodoi{10.1088/2041-8205/788/2/L29}

\bibitem[{Bernardi {et~al.}(2006)Bernardi, Sheth, Nichol, Miller, Schlegel,
  Frieman, Schneider, Subbarao, York, \& Brinkmann}]{Bernardi2006}
Bernardi, M., Sheth, R.~K., Nichol, R.~C., {et~al.} 2006, The Astronomical
  Journal, 131, 2018, \dodoi{10.1086/499770}

\bibitem[{Bertin {et~al.}(2002)Bertin, Ciotti, \& {Del Principe}}]{Bertin2002}
Bertin, G., Ciotti, L., \& {Del Principe}, M. 2002, Astronomy {\&}
  Astrophysics, 386, 149, \dodoi{10.1051/0004-6361:20020248}

\bibitem[{Bezanson {et~al.}(2012)Bezanson, {Van Dokkum}, \&
  Franx}]{Bezanson2012}
Bezanson, R., {Van Dokkum}, P., \& Franx, M. 2012, Astrophysical Journal, 760,
  \dodoi{10.1088/0004-637X/760/1/62}

\bibitem[{Bezanson {et~al.}(2013{\natexlab{a}})Bezanson, van Dokkum, van~de
  Sande, Franx, \& Kriek}]{Bezanson2013a}
Bezanson, R., van Dokkum, P., van~de Sande, J., Franx, M., \& Kriek, M.
  2013{\natexlab{a}}, The Astrophysical Journal, 764, L8,
  \dodoi{10.1088/2041-8205/764/1/L8}

\bibitem[{Bezanson {et~al.}(2009)Bezanson, {Van Dokkum}, Tal, Marchesini,
  Kriek, Franx, \& Coppi}]{Bezanson2009}
Bezanson, R., {Van Dokkum}, P.~G., Tal, T., {et~al.} 2009, Astrophysical
  Journal, 697, 1290, \dodoi{10.1088/0004-637X/697/2/1290}

\bibitem[{Bezanson {et~al.}(2013{\natexlab{b}})Bezanson, van Dokkum, van~de
  Sande, Franx, Leja, \& Kriek}]{Bezanson2013b}
Bezanson, R., van Dokkum, P.~G., van~de Sande, J., {et~al.} 2013{\natexlab{b}},
  The Astrophysical Journal, 779, L21, \dodoi{10.1088/2041-8205/779/2/L21}

\bibitem[{Bezanson {et~al.}(2018)Bezanson, van~der Wel, Pacifici, Noeske,
  Bari{\v{s}}i{\'{c}}, Bell, Brammer, Calhau, Chauke, van Dokkum, Franx,
  Gallazzi, van Houdt, Labb{\'{e}}, Maseda, Mu{\~{n}}os-Mateos, Muzzin, van~de
  Sande, Sobral, Straatman, \& Wu}]{Bezanson2018}
Bezanson, R., van~der Wel, A., Pacifici, C., {et~al.} 2018, The Astrophysical
  Journal, 858, 60, \dodoi{10.3847/1538-4357/aabc55}

\bibitem[{Bruzual \& Charlot(2003)}]{Bruzual2003}
Bruzual, G., \& Charlot, S. 2003, Monthly Notices of the Royal Astronomical
  Society, 344, 1000, \dodoi{10.1046/j.1365-8711.2003.06897.x}

\bibitem[{Cappellari(2017)}]{Cappellari2017}
Cappellari, M. 2017, Monthly Notices of the Royal Astronomical Society, 466,
  798, \dodoi{10.1093/mnras/stw3020}

\bibitem[{Cappellari \& Emsellem(2004)}]{Cappellari2004}
Cappellari, M., \& Emsellem, E. 2004, Publications of the Astronomical Society
  of the Pacific, 116, 138, \dodoi{10.1086/381875}

\bibitem[{Cappellari {et~al.}(2006)Cappellari, Bacon, Bureau, Damen, Davies,
  {De Zeeuw}, Emsellem, Falc{\'{o}}n-Barroso, Krajnovi{\'{e}}, Kuntschner,
  McDermid, Peletier, Sarzi, {Van Den Bosch}, \& {Van De Ven}}]{Cappellari2006}
Cappellari, M., Bacon, R., Bureau, M., {et~al.} 2006, Monthly Notices of the
  Royal Astronomical Society, 366, 1126,
  \dodoi{10.1111/j.1365-2966.2005.09981.x}

\bibitem[{Cappellari {et~al.}(2009)Cappellari, {di Serego Alighieri}, Cimatti,
  Daddi, Renzini, Kurk, Cassata, Dickinson, Franceschini, Mignoli, Pozzetti,
  Rodighiero, Rosati, \& Zamorani}]{Cappellari2009}
Cappellari, M., {di Serego Alighieri}, S., Cimatti, A., {et~al.} 2009, The
  Astrophysical Journal, 704, L34, \dodoi{10.1088/0004-637X/704/1/L34}

\bibitem[{Cappellari {et~al.}(2013{\natexlab{a}})Cappellari, Scott, Alatalo,
  Blitz, Bois, Bournaud, Bureau, Crocker, Davies, Davis, de~Zeeuw, Duc,
  Emsellem, Khochfar, Krajnovi{\'{c}}, Kuntschner, McDermid, Morganti, Naab,
  Oosterloo, Sarzi, Serra, Weijmans, \& Young}]{Cappellari2013a}
Cappellari, M., Scott, N., Alatalo, K., {et~al.} 2013{\natexlab{a}}, Monthly
  Notices of the Royal Astronomical Society, 432, 1709,
  \dodoi{10.1093/mnras/stt562}

\bibitem[{Cappellari {et~al.}(2013{\natexlab{b}})Cappellari, McDermid, Alatalo,
  Blitz, Bois, Bournaud, Bureau, Crocker, Davies, Davis, de~Zeeuw, Duc,
  Emsellem, Khochfar, Krajnovi{\'{c}}, Kuntschner, Morganti, Naab, Oosterloo,
  Sarzi, Scott, Serra, Weijmans, \& Young}]{Cappellari2013b}
Cappellari, M., McDermid, R.~M., Alatalo, K., {et~al.} 2013{\natexlab{b}},
  Monthly Notices of the Royal Astronomical Society, 432, 1862,
  \dodoi{10.1093/mnras/stt644}

\bibitem[{Chabrier(2003)}]{Chabrier2003}
Chabrier, G. 2003, Publications of the Astronomical Society of the Pacific,
  115, 763, \dodoi{10.1086/376392}

\bibitem[{Conroy {et~al.}(2013)Conroy, Dutton, Graves, Mendel, \& {Van
  Dokkum}}]{Conroy2013}
Conroy, C., Dutton, A.~A., Graves, G.~J., Mendel, J.~T., \& {Van Dokkum}, P.~G.
  2013, Astrophysical Journal Letters, 776, \dodoi{10.1088/2041-8205/776/2/L26}

\bibitem[{Conroy \& van Dokkum(2012)}]{Conroy2012}
Conroy, C., \& van Dokkum, P.~G. 2012, The Astrophysical Journal, 760, 71,
  \dodoi{10.1088/0004-637X/760/1/71}

\bibitem[{Conroy {et~al.}(2017)Conroy, van Dokkum, \& Villaume}]{Conroy2017}
Conroy, C., van Dokkum, P.~G., \& Villaume, A. 2017, The Astrophysical Journal,
  837, 166, \dodoi{10.3847/1538-4357/aa6190}

\bibitem[{Cutler {et~al.}(2022)Cutler, Whitaker, Mowla, Brammer, van~der Wel,
  Marchesini, van Dokkum, Momcheva, Song, Akhshik, Nelson, Bezanson, Franx,
  Kriek, Lange-Vagle, Leja, MacKenty, Muzzin, \& Shipley}]{Cutler2022}
Cutler, S.~E., Whitaker, K.~E., Mowla, L.~A., {et~al.} 2022, The Astrophysical
  Journal, 925, 34, \dodoi{10.3847/1538-4357/ac341c}

\bibitem[{Djorgovski \& Davis(1987)}]{Djorgovski1987}
Djorgovski, S., \& Davis, M. 1987, The Astrophysical Journal, 313, 59,
  \dodoi{10.1086/164948}

\bibitem[{Donnari {et~al.}(2021)Donnari, Pillepich, Nelson, Marinacci,
  Vogelsberger, \& Hernquist}]{Donnari2021}
Donnari, M., Pillepich, A., Nelson, D., {et~al.} 2021, Monthly Notices of the
  Royal Astronomical Society, 506, 4760, \dodoi{10.1093/mnras/stab1950}

\bibitem[{Dressler(1987)}]{Dressler1987}
Dressler, A. 1987, The Astrophysical Journal, 317, 1, \dodoi{10.1086/165251}

\bibitem[{Esdaile {et~al.}(2021)Esdaile, Glazebrook, Labb{\'{e}}, Taylor,
  Schreiber, Nanayakkara, Kacprzak, Oesch, Tran, Papovich, Spitler, \&
  Straatman}]{Esdaile2021}
Esdaile, J., Glazebrook, K., Labb{\'{e}}, I., {et~al.} 2021, The Astrophysical
  Journal, 908, L35, \dodoi{10.3847/2041-8213/abe11e}

\bibitem[{Faber \& Jackson(1976)}]{Faber1976}
Faber, S.~M., \& Jackson, R.~E. 1976, The Astrophysical Journal, 204, 668,
  \dodoi{10.1086/154215}

\bibitem[{Forrest {et~al.}(2017)Forrest, Tran, Broussard, Allen, Apfel, Cowley,
  Glazebrook, Kacprzak, Labb{\'{e}}, Nanayakkara, Papovich, Quadri, Spitler,
  Straatman, \& Tomczak}]{Forrest2017}
Forrest, B., Tran, K.-V.~H., Broussard, A., {et~al.} 2017, The Astrophysical
  Journal, 838, L12, \dodoi{10.3847/2041-8213/aa653b}

\bibitem[{Forrest {et~al.}(2020{\natexlab{a}})Forrest, Annunziatella, Wilson,
  Marchesini, Muzzin, Cooper, Marsan, McConachie, Chan, Gomez, Kado-Fong,
  Barbera, Labb{\'{e}}, Lange-Vagle, Nantais, Nonino, Pe{\~{n}}a, Saracco,
  Stefanon, \& van~der Burg}]{Forrest2020a}
Forrest, B., Annunziatella, M., Wilson, G., {et~al.} 2020{\natexlab{a}}, The
  Astrophysical Journal, 890, L1, \dodoi{10.3847/2041-8213/ab5b9f}

\bibitem[{Forrest {et~al.}(2020{\natexlab{b}})Forrest, Marsan, Annunziatella,
  Wilson, Muzzin, Marchesini, Cooper, Chan, McConachie, Gomez, Kado-Fong, {La
  Barbera}, Lange-Vagle, Nantais, Nonino, Saracco, Stefanon, \& van~der
  Burg}]{Forrest2020b}
Forrest, B., Marsan, Z.~C., Annunziatella, M., {et~al.} 2020{\natexlab{b}}, The
  Astrophysical Journal, 903, 47, \dodoi{10.3847/1538-4357/abb819}

\bibitem[{Gargiulo {et~al.}(2016)Gargiulo, Saracco, Tamburri, Lonoce, \&
  Ciocca}]{Gargiulo2016}
Gargiulo, A., Saracco, P., Tamburri, S., Lonoce, I., \& Ciocca, F. 2016,
  Astronomy and Astrophysics, 592, 1, \dodoi{10.1051/0004-6361/201526563}

\bibitem[{Gebhardt {et~al.}(2000)Gebhardt, Bender, Bower, Dressler, Faber,
  Filippenko, Green, Grillmair, Ho, Kormendy, Lauer, Magorrian, Pinkney,
  Richstone, \& Tremaine}]{Gebhardt2000}
Gebhardt, K., Bender, R., Bower, G., {et~al.} 2000, The Astrophysical Journal,
  539, L13, \dodoi{10.1086/312840}

\bibitem[{Glazebrook {et~al.}(2017)Glazebrook, Schreiber, Labb{\'{e}},
  Nanayakkara, Kacprzak, Oesch, Papovich, Spitler, Straatman, Tran, \&
  Yuan}]{Glazebrook2017}
Glazebrook, K., Schreiber, C., Labb{\'{e}}, I., {et~al.} 2017, Nature
  Publishing Group, 544, 71, \dodoi{10.1038/nature21680}

\bibitem[{Haussler {et~al.}(2007)Haussler, McIntosh, Barden, Bell, Rix, Borch,
  Beckwith, Caldwell, Heymans, Jahnke, Jogee, Koposov, Meisenheimer, Sanchez,
  Somerville, Wisotzki, \& Wolf}]{Haussler2007}
Haussler, B., McIntosh, D.~H., Barden, M., {et~al.} 2007, The Astrophysical
  Journal Supplement Series, 172, 615, \dodoi{10.1086/518836}

\bibitem[{H{\"{a}}u{\ss}ler {et~al.}(2013)H{\"{a}}u{\ss}ler, Bamford, Vika,
  Rojas, Barden, Kelvin, Alpaslan, Robotham, Driver, Baldry, Brough, Hopkins,
  Liske, Nichol, Popescu, \& Tuffs}]{Haussler2013}
H{\"{a}}u{\ss}ler, B., Bamford, S.~P., Vika, M., {et~al.} 2013, Monthly Notices
  of the Royal Astronomical Society, 430, 330, \dodoi{10.1093/mnras/sts633}

\bibitem[{Hill {et~al.}(2016)Hill, Muzzin, Franx, \& van~de Sande}]{Hill2016}
Hill, A.~R., Muzzin, A., Franx, M., \& van~de Sande, J. 2016, The Astrophysical
  Journal, 819, 0, \dodoi{10.3847/0004-637X/819/1/74}

\bibitem[{Hopkins(2018)}]{Hopkins2018a}
Hopkins, A.~M. 2018, Publications of the Astronomical Society of Australia, 35,
  e039, \dodoi{10.1017/pasa.2018.29}

\bibitem[{Hopkins {et~al.}(2009)Hopkins, Lauer, Cox, Hernquist, \&
  Kormendy}]{Hopkins2009}
Hopkins, P.~F., Lauer, T.~R., Cox, T.~J., Hernquist, L., \& Kormendy, J. 2009,
  Astrophysical Journal, Supplement Series, 181, 486,
  \dodoi{10.1088/0067-0049/181/2/486}

\bibitem[{Hunter(2007)}]{Hunter2007}
Hunter, J.~D. 2007, Computing in Science and Engineering,
  \dodoi{10.1109/MCSE.2007.55}

\bibitem[{Husser {et~al.}(2013)Husser, {Wende-von Berg}, Dreizler, Homeier,
  Reiners, Barman, \& Hauschildt}]{Husser2013}
Husser, T.-O., {Wende-von Berg}, S., Dreizler, S., {et~al.} 2013, Astronomy
  {\&} Astrophysics, 553, A6, \dodoi{10.1051/0004-6361/201219058}

\bibitem[{Jarvis {et~al.}(2013)Jarvis, Bonfield, Bruce, Geach, McAlpine,
  McLure, Gonz{\'{a}}lez-Solares, Irwin, Lewis, Yoldas, Andreon, Cross,
  Emerson, Dalton, Dunlop, Hodgkin, Le, Karouzos, Meisenheimer, Oliver,
  Rawlings, Simpson, Smail, Smith, Sullivan, Sutherland, White, \&
  Zwart}]{Jarvis2013}
Jarvis, M.~J., Bonfield, D.~G., Bruce, V.~A., {et~al.} 2013, Monthly Notices of
  the Royal Astronomical Society, 428, 1281, \dodoi{10.1093/mnras/sts118}

\bibitem[{Jorgensen {et~al.}(1995)Jorgensen, Franx, \&
  Kjaergaard}]{Jorgensen1995}
Jorgensen, I., Franx, M., \& Kjaergaard, P. 1995, Monthly Notices of the Royal
  Astronomical Society, 276, 1341, \dodoi{10.1093/mnras/276.4.1341}

\bibitem[{Kausch {et~al.}(2015)Kausch, Noll, Smette, Kimeswenger, Barden,
  Szyszka, Jones, Sana, Horst, \& Kerber}]{Kausch2015}
Kausch, W., Noll, S., Smette, A., {et~al.} 2015, Astronomy and Astrophysics,
  576, \dodoi{10.1051/0004-6361/201423909}

\bibitem[{K{\"{o}}ppen {et~al.}(2007)K{\"{o}}ppen, Weidner, \&
  Kroupa}]{Koppen2007}
K{\"{o}}ppen, J., Weidner, C., \& Kroupa, P. 2007, Monthly Notices of the Royal
  Astronomical Society, 375, 673, \dodoi{10.1111/j.1365-2966.2006.11328.x}

\bibitem[{Kormendy \& Ho(2013)}]{Kormendy2013}
Kormendy, J., \& Ho, L.~C. 2013, Annual Review of Astronomy and Astrophysics,
  51, 511, \dodoi{10.1146/annurev-astro-082708-101811}

\bibitem[{Kriek {et~al.}(2009)Kriek, van Dokkum, Labb{\'{e}}, Franx,
  Illingworth, Marchesini, \& Quadri}]{Kriek2009}
Kriek, M., van Dokkum, P.~G., Labb{\'{e}}, I., {et~al.} 2009, The Astrophysical
  Journal, 700, 221, \dodoi{10.1088/0004-637X/700/1/221}

\bibitem[{Kroupa {et~al.}(2013)Kroupa, Weidner, Pflamm-Altenburg, Thies,
  Dabringhausen, Marks, \& Maschberger}]{Kroupa2013}
Kroupa, P., Weidner, C., Pflamm-Altenburg, J., {et~al.} 2013, Planets, Stars
  and Stellar Systems: Volume 5: Galactic Structure and Stellar Populations, 5,
  115, \dodoi{10.1007/978-94-007-5612-0_4}

\bibitem[{{La Barbera} {et~al.}(2019){La Barbera}, Vazdekis, Ferreras,
  Pasquali, Prieto, Mart{\'{i}}n-Navarro, Aguado, {De Carvalho}, Rembold,
  Falc{\'{o}}n-Barroso, \& {Van De Ven}}]{LaBarbera2019}
{La Barbera}, F., Vazdekis, A., Ferreras, I., {et~al.} 2019, Monthly Notices of
  the Royal Astronomical Society, 489, 4090, \dodoi{10.1093/mnras/stz2192}

\bibitem[{Lacy {et~al.}(2021)Lacy, Surace, Farrah, Nyland, Afonso, Brandt,
  Clements, Lagos, Maraston, Pforr, Sajina, Sako, Vaccari, Wilson, Ballantyne,
  Barkhouse, Brunner, Cane, Clarke, Cooper, Cooray, Covone, D'Andrea, Evrard,
  Ferguson, Frieman, Gonzalez-Perez, Gupta, Hatziminaoglou, Huang, Jagannathan,
  Jarvis, Jones, Kimball, Lidman, Lubin, Marchetti, Martini, Mcmahon, Mei,
  Messias, Murphy, Newman, Nichol, Norris, Oliver, Perez-Fournon, Peters,
  Pierre, Polisensky, Richards, Ridgway, R{\"{o}}ttgering, Seymour, Shirley,
  Somerville, Strauss, Suntzeff, Thorman, {Van Kampen}, Verma, Wechsler, \&
  Wood-Vasey}]{Lacy2021}
Lacy, M., Surace, J.~A., Farrah, D., {et~al.} 2021, Monthly Notices of the
  Royal Astronomical Society, 501, 892, \dodoi{10.1093/mnras/staa3714}

\bibitem[{L{\"{a}}sker {et~al.}(2013)L{\"{a}}sker, van~den Bosch, van~de Ven,
  Ferreras, {La Barbera}, Vazdekis, \& Falc{\'{o}}n-Barroso}]{Lasker2013}
L{\"{a}}sker, R., van~den Bosch, R.~C., van~de Ven, G., {et~al.} 2013, Monthly
  Notices of the Royal Astronomical Society: Letters, 434,
  \dodoi{10.1093/mnrasl/slt070}

\bibitem[{Leja {et~al.}(2018)Leja, Johnson, Conroy, \& van Dokkum}]{Leja2018}
Leja, J., Johnson, B.~D., Conroy, C., \& van Dokkum, P. 2018, The Astrophysical
  Journal, 854, 62, \dodoi{10.3847/1538-4357/aaa8db}

\bibitem[{Leja {et~al.}(2017)Leja, Johnson, Conroy, van Dokkum, \&
  Byler}]{Leja2017}
Leja, J., Johnson, B.~D., Conroy, C., van Dokkum, P.~G., \& Byler, N. 2017, The
  Astrophysical Journal, 837, 170, \dodoi{10.3847/1538-4357/aa5ffe}

\bibitem[{Leja {et~al.}(2019)Leja, Johnson, Conroy, van Dokkum, Speagle,
  Brammer, Momcheva, Skelton, Whitaker, Franx, \& Nelson}]{Leja2019}
Leja, J., Johnson, B.~D., Conroy, C., {et~al.} 2019, The Astrophysical Journal,
  877, 140, \dodoi{10.3847/1538-4357/ab1d5a}

\bibitem[{Lian {et~al.}(2018)Lian, Thomas, \& Maraston}]{Lian2018}
Lian, J., Thomas, D., \& Maraston, C. 2018, Monthly Notices of the Royal
  Astronomical Society, 39, 6027, \dodoi{10.1093/mnras/sty2506}

\bibitem[{Lustig {et~al.}(2021)Lustig, Strazzullo, D'Eugenio, Daddi, Pannella,
  Renzini, Cimatti, Gobat, Jin, Mohr, \& Onodera}]{Lustig2020}
Lustig, P., Strazzullo, V., D'Eugenio, C., {et~al.} 2021, Monthly Notices of
  the Royal Astronomical Society, 501, 2659, \dodoi{10.1093/mnras/staa3766}

\bibitem[{Lustig {et~al.}(2022)Lustig, Strazzullo, Remus, D'Eugenio, Daddi,
  Burkert, {De Lucia}, Delvecchio, Dolag, Fontanot, Gobat, Mohr, Onodera,
  Pannella, Pillepich, \& Renzini}]{Lustig2022}
Lustig, P., Strazzullo, V., Remus, R.-S., {et~al.} 2022, 19, 1.
\newblock \doarXiv{2201.09068}

\bibitem[{Marsan {et~al.}(2017)Marsan, Marchesini, Brammer, Geier, Kado-Fong,
  Labb{\'{e}}, Muzzin, \& Stefanon}]{Marsan2017}
Marsan, Z.~C., Marchesini, D., Brammer, G.~B., {et~al.} 2017, The Astrophysical
  Journal, 842, 21, \dodoi{10.3847/1538-4357/aa7206}

\bibitem[{Marsan {et~al.}(2015)Marsan, Marchesini, Brammer, Stefanon, Muzzin,
  Fern{\'{a}}ndez-soto, Geier, Hainline, Intema, Karim, \&
  Labb{\'{e}}}]{Marsan2015}
---. 2015, The Astrophysical Journal, 801, 133,
  \dodoi{10.1088/0004-637X/801/2/133}

\bibitem[{Marsan {et~al.}(2019)Marsan, Marchesini, Muzzin, Brammer, Bezanson,
  Franx, Labb{\'{e}}, Lundgren, Rudnick, Stefanon, \& Dokkum}]{Marsan2019}
Marsan, Z.~C., Marchesini, D., Muzzin, A., {et~al.} 2019, The Astrophysical
  Journal, 871, 201, \dodoi{10.3847/1538-4357/aaf808}

\bibitem[{Marsan {et~al.}(2022)Marsan, Muzzin, Marchesini, Stefanon, Martis,
  Annunziatella, Chan, Cooper, Forrest, Gomez, McConachie, \&
  Wilson}]{Marsan2022}
Marsan, Z.~C., Muzzin, A., Marchesini, D., {et~al.} 2022, The Astrophysical
  Journal, 924, 25, \dodoi{10.3847/1538-4357/ac312a}

\bibitem[{Mart{\'{i}}n-Navarro {et~al.}(2015)Mart{\'{i}}n-Navarro, Vazdekis,
  {La Barbera}, Falc{\'{o}}n-Barroso, Lyubenova, {Van De Ven}, Ferreras,
  S{\'{a}}nchez, Trager, Garc{\'{i}}a-Benito, Mast, Mendoza,
  S{\'{a}}nchez-Bl{\'{a}}zquez, Delgado, \& Walcher}]{Martin-Navarro2015}
Mart{\'{i}}n-Navarro, I., Vazdekis, A., {La Barbera}, F., {et~al.} 2015,
  Astrophysical Journal Letters, 806, L31, \dodoi{10.1088/2041-8205/806/2/L31}

\bibitem[{Mauduit {et~al.}(2012)Mauduit, Lacy, Farrah, Surace, Jarvis, Oliver,
  Maraston, Vaccari, Marchetti, Zeimann, Gonz{\'{a}}les-Solares, Pforr, Petric,
  Henriques, Thomas, Afonso, Rettura, Wilson, Falder, Geach, Huynh, Norris,
  Seymour, Richards, Stanford, Alexander, Becker, Best, Bizzocchi, Bonfield,
  Castro, Cava, Chapman, Christopher, Clements, Covone, Dubois, Dunlop, Dyke,
  Edge, Ferguson, Foucaud, Franceschini, Gal, Grant, Grossi, Hatziminaoglou,
  Hickey, Hodge, Huang, Ivison, Kim, LeFevre, Lehnert, Lonsdale, Lubin, McLure,
  Messias, Mart{\'{i}}nez-Sansigre, Mortier, Nielsen, Ouchi, Parish,
  Perez-Fournon, Pierre, Rawlings, Readhead, Ridgway, Rigopoulou, Romer,
  Rosebloom, Rottgering, Rowan-Robinson, Sajina, Simpson, Smail, Squires,
  Stevens, Taylor, Trichas, Urrutia, van Kampen, Verma, \& Xu}]{Mauduit2012}
Mauduit, J.-C., Lacy, M., Farrah, D., {et~al.} 2012, Publications of the
  Astronomical Society of the Pacific, 124, 1135, \dodoi{10.1086/668290}

\bibitem[{McCracken {et~al.}(2012)McCracken, Milvang-Jensen, Dunlop, Franx,
  Fynbo, {Le F{\`{e}}vre}, Holt, Caputi, Goranova, Buitrago, Emerson,
  Freudling, Hudelot, L{\'{o}}pez-Sanjuan, Magnard, Mellier, M{\o}ller,
  Nilsson, Sutherland, Tasca, \& Zabl}]{McCracken2012}
McCracken, H.~J., Milvang-Jensen, B., Dunlop, J., {et~al.} 2012, Astronomy {\&}
  Astrophysics, 544, A156, \dodoi{10.1051/0004-6361/201219507}

\bibitem[{McLean {et~al.}(2010)McLean, Steidel, Epps, Matthews, Adkins,
  Konidaris, Weber, Aliado, Brims, Canfield, Cromer, Fucik, Kulas, Mace,
  Magnone, Rodriguez, Wang, \& Weiss}]{McLean2010}
McLean, I.~S., Steidel, C.~C., Epps, H., {et~al.} 2010in , 77351E.
\newblock
  \url{http://proceedings.spiedigitallibrary.org/proceeding.aspx?doi=10.1117/12.856715}

\bibitem[{McLean {et~al.}(2012)McLean, Steidel, Epps, Konidaris, Matthews,
  Adkins, Aliado, Brims, Canfield, Cromer, Fucik, Kulas, Mace, Magnone,
  Rodriguez, Rudie, Trainor, Wang, Weber, \& Weiss}]{McLean2012}
McLean, I.~S., Steidel, C.~C., Epps, H.~W., {et~al.} 2012in .
\newblock
  \url{http://proceedings.spiedigitallibrary.org/proceeding.aspx?doi=10.1117/12.924794}

\bibitem[{Mendel {et~al.}(2020)Mendel, Beifiori, Saglia, Bender, Brammer, Chan,
  {F{\"{o}}rster Schreiber}, Fossati, Galametz, Momcheva, Nelson, Wilman, \&
  Wuyts}]{Mendel2020}
Mendel, J.~T., Beifiori, A., Saglia, R.~P., {et~al.} 2020, The Astrophysical
  Journal, 899, 87, \dodoi{10.3847/1538-4357/ab9ffc}

\bibitem[{Mobasher {et~al.}(2015)Mobasher, Dahlen, Ferguson, Acquaviva, Barro,
  Finkelstein, Fontana, Gruetzbauch, Johnson, Lu, Papovich, Pforr, Salvato,
  Somerville, Wiklind, Wuyts, Ashby, Bell, Conselice, Dickinson, Faber, Fazio,
  Finlator, Galametz, Gawiser, Giavalisco, Grazian, Grogin, Guo, Hathi,
  Kocevski, Koekemoer, Koo, Newman, Reddy, Santini, \& Wechsler}]{Mobasher2015}
Mobasher, B., Dahlen, T., Ferguson, H.~C., {et~al.} 2015, Astrophysical
  Journal, 808, \dodoi{10.1088/0004-637X/808/1/101}

\bibitem[{Mowla {et~al.}(2019)Mowla, van Dokkum, Brammer, Momcheva, van~der
  Wel, Whitaker, Nelson, Bezanson, Muzzin, Franx, MacKenty, Leja, Kriek, \&
  Marchesini}]{Mowla2019a}
Mowla, L.~A., van Dokkum, P., Brammer, G.~B., {et~al.} 2019, The Astrophysical
  Journal, 880, 57, \dodoi{10.3847/1538-4357/ab290a}

\bibitem[{Muzzin {et~al.}(2009)Muzzin, Wilson, Yee, Hoekstra, Gilbank, Surace,
  Lacy, Blindert, Majumdar, Demarco, Gardner, Gladders, \&
  Lonsdale}]{Muzzin2009}
Muzzin, A., Wilson, G., Yee, H. K.~C., {et~al.} 2009, The Astrophysical
  Journal, 698, 1934, \dodoi{10.1088/0004-637X/698/2/1934}

\bibitem[{Muzzin {et~al.}(2013{\natexlab{a}})Muzzin, Marchesini, Stefanon,
  Franx, McCracken, Milvang-Jensen, Dunlop, Fynbo, Brammer, Labb{\'{e}}, \&
  {Van Dokkum}}]{Muzzin2013}
Muzzin, A., Marchesini, D., Stefanon, M., {et~al.} 2013{\natexlab{a}},
  Astrophysical Journal, 777, \dodoi{10.1088/0004-637X/777/1/18}

\bibitem[{Muzzin {et~al.}(2013{\natexlab{b}})Muzzin, Marchesini, Stefanon,
  Franx, Milvang-Jensen, Dunlop, Fynbo, Brammer, Labb{\'{e}}, \& van
  Dokkum}]{Muzzin2013a}
---. 2013{\natexlab{b}}, The Astrophysical Journal Supplement Series, 206, 8,
  \dodoi{10.1088/0067-0049/206/1/8}

\bibitem[{Nedkova {et~al.}(2021)Nedkova, H{\"{a}}u{\ss}ler, Marchesini,
  Dimauro, Brammer, Eigenthaler, Feinstein, Ferguson, Huertas-Company,
  Johnston, Kado-Fong, Kartaltepe, Labb{\'{e}}, Lange-Vagle, Martis, McGrath,
  Muzzin, Oesch, Ordenes-Brice{\~{n}}o, Puzia, Shipley, Simmons, Skelton,
  Stefanon, van der Wel, \& Whitaker}]{Nedkova2021}
Nedkova, K.~V., H{\"{a}}u{\ss}ler, B., Marchesini, D., {et~al.} 2021, Monthly
  Notices of the Royal Astronomical Society, 506, 928,
  \dodoi{10.1093/mnras/stab1744}

\bibitem[{Newman {et~al.}(2018)Newman, Belli, Ellis, \& Patel}]{Newman2018b}
Newman, A.~B., Belli, S., Ellis, R.~S., \& Patel, S.~G. 2018, The Astrophysical
  Journal, 862, 126, \dodoi{10.3847/1538-4357/aacd4f}

\bibitem[{Newman {et~al.}(2012)Newman, Ellis, Bundy, \& Treu}]{Newman2012}
Newman, A.~B., Ellis, R.~S., Bundy, K., \& Treu, T. 2012, Astrophysical
  Journal, 746, \dodoi{10.1088/0004-637X/746/2/162}

\bibitem[{Newman {et~al.}(2010)Newman, Ellis, Treu, \& Bundy}]{Newman2010}
Newman, A.~B., Ellis, R.~S., Treu, T., \& Bundy, K. 2010, The Astrophysical
  Journal, 717, L103, \dodoi{10.1088/2041-8205/717/2/L103}

\bibitem[{Oke \& Gunn(1983)}]{Oke1983}
Oke, J.~B., \& Gunn, J.~E. 1983, The Astrophysical Journal, 266, 713,
  \dodoi{10.1086/160817}

\bibitem[{Oliphant \& Millma(2006)}]{Oliphant2006}
Oliphant, T., \& Millma, J.~k. 2006, {A guide to NumPy},
  \dodoi{DOI:10.1109/MCSE.2007.58}

\bibitem[{Peng {et~al.}(2002)Peng, Ho, Impey, \& Rix}]{Peng2002}
Peng, C.~Y., Ho, L.~C., Impey, C.~D., \& Rix, H.-W. 2002, The Astronomical
  Journal, 124, 266, \dodoi{10.1086/340952}

\bibitem[{Peng {et~al.}(2010)Peng, Ho, Impey, \& Rix}]{Peng2010}
Peng, C.~Y., Ho, L.~C., Impey, C.~D., \& Rix, H.~W. 2010, Astronomical Journal,
  139, 2097, \dodoi{10.1088/0004-6256/139/6/2097}

\bibitem[{P{\'{e}}rez \& Granger(2007)}]{Perez2007}
P{\'{e}}rez, F., \& Granger, B.~E. 2007, Computing in Science and Engineering,
  \dodoi{10.1109/MCSE.2007.53}

\bibitem[{Posacki {et~al.}(2015)Posacki, Cappellari, Treu, Pellegrini, \&
  Ciotti}]{Posacki2015}
Posacki, S., Cappellari, M., Treu, T., Pellegrini, S., \& Ciotti, L. 2015,
  Monthly Notices of the Royal Astronomical Society, 446, 493,
  \dodoi{10.1093/mnras/stu2098}

\bibitem[{Price-Whelan {et~al.}(2018)Price-Whelan, Sipőcz, G{\"{u}}nther, Lim,
  Crawford, Conseil, Shupe, Craig, Dencheva, Ginsburg, VanderPlas, Bradley,
  P{\'{e}}rez-Su{\'{a}}rez, de~Val-Borro, Aldcroft, Cruz, Robitaille, Tollerud,
  Ardelean, Babej, Bach, Bachetti, Bakanov, Bamford, Barentsen, Barmby,
  Baumbach, Berry, Biscani, Boquien, Bostroem, Bouma, Brammer, Bray,
  Breytenbach, Buddelmeijer, Burke, Calderone, Rodr{\'{i}}guez, Cara, Cardoso,
  Cheedella, Copin, Corrales, Crichton, D'Avella, Deil, Depagne, Dietrich,
  Donath, Droettboom, Earl, Erben, Fabbro, Ferreira, Finethy, Fox, Garrison,
  Gibbons, Goldstein, Gommers, Greco, Greenfield, Groener, Grollier, Hagen,
  Hirst, Homeier, Horton, Hosseinzadeh, Hu, Hunkeler, Ivezi{\'{c}}, Jain,
  Jenness, Kanarek, Kendrew, Kern, Kerzendorf, Khvalko, King, Kirkby, Kulkarni,
  Kumar, Lee, Lenz, Littlefair, Ma, Macleod, Mastropietro, McCully, Montagnac,
  Morris, Mueller, Mumford, Muna, Murphy, Nelson, Nguyen, Ninan, N{\"{o}}the,
  Ogaz, Oh, Parejko, Parley, Pascual, Patil, Patil, Plunkett, Prochaska,
  Rastogi, Janga, Sabater, Sakurikar, Seifert, Sherbert, Sherwood-Taylor, Shih,
  Sick, Silbiger, Singanamalla, Singer, Sladen, Sooley, Sornarajah, Streicher,
  Teuben, Thomas, Tremblay, Turner, Terr{\'{o}}n, van Kerkwijk, de~la Vega,
  Watkins, Weaver, Whitmore, Woillez, \& Zabalza}]{Astropy2018}
Price-Whelan, A.~M., Sipőcz, B.~M., G{\"{u}}nther, H.~M., {et~al.} 2018, The
  Astronomical Journal, 156, 123, \dodoi{10.3847/1538-3881/aabc4f}

\bibitem[{Prochaska {et~al.}(2020)Prochaska, Hennawi, Westfall, Cooke, Wang,
  Hsyu, Davies, Farina, \& Pelliccia}]{Prochaska2020}
Prochaska, J., Hennawi, J., Westfall, K., {et~al.} 2020, Journal of Open Source
  Software, 5, 2308, \dodoi{10.21105/joss.02308}

\bibitem[{Ribeiro {et~al.}(2016)Ribeiro, {Le F{\`{e}}vre}, Tasca, Lemaux,
  Cassata, Garilli, Maccagni, Zamorani, Zucca, Amor{\'{i}}n, Bardelli, Fontana,
  Giavalisco, Hathi, Koekemoer, Pforr, Tresse, \& Dunlop}]{Ribeiro2016}
Ribeiro, B., {Le F{\`{e}}vre}, O., Tasca, L. A.~M., {et~al.} 2016, Astronomy
  {\&} Astrophysics, 593, A22, \dodoi{10.1051/0004-6361/201628249}

\bibitem[{Robitaille {et~al.}(2013)Robitaille, Tollerud, Greenfield,
  Droettboom, Bray, Aldcroft, Davis, Ginsburg, Price-Whelan, Kerzendorf,
  Conley, Crighton, Barbary, Muna, Ferguson, Grollier, Parikh, Nair,
  G{\"{u}}nther, Deil, Woillez, Conseil, Kramer, Turner, Singer, Fox, Weaver,
  Zabalza, Edwards, {Azalee Bostroem}, Burke, Casey, Crawford, Dencheva, Ely,
  Jenness, Labrie, Lim, Pierfederici, Pontzen, Ptak, Refsdal, Servillat, \&
  Streicher}]{Astropy2013}
Robitaille, T.~P., Tollerud, E.~J., Greenfield, P., {et~al.} 2013, Astronomy
  {\&} Astrophysics, 558, A33, \dodoi{10.1051/0004-6361/201322068}

\bibitem[{Salmon {et~al.}(2015)Salmon, Papovich, Finkelstein, Tilvi, Finlator,
  Behroozi, Dahlen, Dav{\'{e}}, Dekel, Dickinson, Ferguson, Giavalisco, Long,
  Lu, Mobasher, Reddy, Somerville, \& Wechsler}]{Salmon2015}
Salmon, B., Papovich, C., Finkelstein, S.~L., {et~al.} 2015, The Astrophysical
  Journal, 799, 183, \dodoi{10.1088/0004-637X/799/2/183}

\bibitem[{Salpeter(1955)}]{Salpeter1955}
Salpeter, E.~E. 1955, The Astrophysical Journal, 121, 161

\bibitem[{S{\'{a}}nchez-Bl{\'{a}}zquez
  {et~al.}(2006)S{\'{a}}nchez-Bl{\'{a}}zquez, Peletier, Jim{\'{e}}nez-Vicente,
  Cardiel, Cenarro, Falc{\'{o}}n-Barroso, Gorgas, Selam, \&
  Vazdekis}]{Sanchez-Blazquez2006}
S{\'{a}}nchez-Bl{\'{a}}zquez, P., Peletier, R.~F., Jim{\'{e}}nez-Vicente, J.,
  {et~al.} 2006, Monthly Notices of the Royal Astronomical Society, 371, 703,
  \dodoi{10.1111/j.1365-2966.2006.10699.x}

\bibitem[{Sanders {et~al.}(2021)Sanders, Shapley, Jones, Reddy, Kriek, Siana,
  Coil, Mobasher, Shivaei, Dav{\'{e}}, Azadi, Price, Leung, Freeman, Fetherolf,
  de~Groot, Zick, \& Barro}]{Sanders2021}
Sanders, R.~L., Shapley, A.~E., Jones, T., {et~al.} 2021, The Astrophysical
  Journal, 914, 19, \dodoi{10.3847/1538-4357/abf4c1}

\bibitem[{Saracco {et~al.}(2019)Saracco, {La Barbera}, Gargiulo, Mannucci,
  Marchesini, Nonino, \& Ciliegi}]{Saracco2019}
Saracco, P., {La Barbera}, F., Gargiulo, A., {et~al.} 2019, Monthly Notices of
  the Royal Astronomical Society, 484, 2281, \dodoi{10.1093/mnras/sty3509}

\bibitem[{Saracco {et~al.}(2020)Saracco, Marchesini, Barbera, Gargiulo,
  Annunziatella, Forrest, {Lange Vagle}, Marsan, Muzzin, Stefanon, \&
  Wilson}]{Saracco2020}
Saracco, P., Marchesini, D., Barbera, F.~L., {et~al.} 2020, The Astrophysical
  Journal, 905, 40, \dodoi{10.3847/1538-4357/abc7c4}

\bibitem[{Saulder {et~al.}(2015)Saulder, {Van Den Bosch}, \&
  Mieske}]{Saulder2015}
Saulder, C., {Van Den Bosch}, R.~C., \& Mieske, S. 2015, Astronomy and
  Astrophysics, 578, 1, \dodoi{10.1051/0004-6361/201425472}

\bibitem[{Schreiber {et~al.}(2018{\natexlab{a}})Schreiber, Glazebrook,
  Nanayakkara, Kacprzak, Labb{\'{e}}, Oesch, Yuan, Tran, Papovich, Spitler, \&
  Straatman}]{Schreiber2018b}
Schreiber, C., Glazebrook, K., Nanayakkara, T., {et~al.} 2018{\natexlab{a}},
  Astronomy {\&} Astrophysics, 618, A85, \dodoi{10.1051/0004-6361/201833070}

\bibitem[{Schreiber {et~al.}(2018{\natexlab{b}})Schreiber, Labb{\'{e}},
  Glazebrook, Bekiaris, Papovich, Costa, Elbaz, Kacprzak, Nanayakkara, Oesch,
  Pannella, Spitler, Straatman, Tran, \& Wang}]{Schreiber2018a}
Schreiber, C., Labb{\'{e}}, I., Glazebrook, K., {et~al.} 2018{\natexlab{b}},
  Astronomy {\&} Astrophysics, 611, A22, \dodoi{10.1051/0004-6361/201731917}

\bibitem[{Sherman {et~al.}(2019)Sherman, Jogee, Florez, Stevans,
  Kawinwanichakij, Wold, Finkelstein, Papovich, Acquaviva, Ciardullo, Gronwall,
  \& Escalante}]{Sherman2019}
Sherman, S., Jogee, S., Florez, J., {et~al.} 2019, Monthly Notices of the Royal
  Astronomical Society, 19, 1, \dodoi{10.1093/mnras/stz3229}

\bibitem[{Shu {et~al.}(2012)Shu, Bolton, Schlegel, Dawson, Wake, Brownstein,
  Brinkmann, \& Weaver}]{Shu2012}
Shu, Y., Bolton, A.~S., Schlegel, D.~J., {et~al.} 2012, Astronomical Journal,
  143, \dodoi{10.1088/0004-6256/143/4/90}

\bibitem[{Smette {et~al.}(2015)Smette, Sana, Noll, Horst, Kausch, Kimeswenger,
  Barden, Szyszka, Jones, Gallenne, Vinther, Ballester, \& Taylor}]{Smette2015}
Smette, A., Sana, H., Noll, S., {et~al.} 2015, Astronomy {\&} Astrophysics,
  576, A77, \dodoi{10.1051/0004-6361/201423932}

\bibitem[{Stark {et~al.}(2013)Stark, Schenker, Ellis, Robertson, McLure, \&
  Dunlop}]{Stark2013}
Stark, D.~P., Schenker, M.~a., Ellis, R., {et~al.} 2013, The Astrophysical
  Journal, 763, 129, \dodoi{10.1088/0004-637X/763/2/129}

\bibitem[{Straatman {et~al.}(2015{\natexlab{a}})Straatman, Labb{\'{e}},
  Spitler, Glazebrook, Tomczak, Allen, Brammer, Cowley, Dokkum, Kacprzak,
  Kawinwanichakij, Mehrtens, Nanayakkara, Papovich, Persson, Quadri, Rees,
  Tilvi, Tran, \& Whitaker}]{Straatman2015}
Straatman, C.~M., Labb{\'{e}}, I., Spitler, L.~R., {et~al.} 2015{\natexlab{a}},
  Astrophysical Journal Letters, 808, L29, \dodoi{10.1088/2041-8205/808/1/L29}

\bibitem[{Straatman {et~al.}(2014)Straatman, Labb{\'{e}}, Spitler, Allen,
  Altieri, Brammer, Dickinson, van Dokkum, Inami, Glazebrook, Kacprzak,
  Kawinwanichakij, Kelson, McCarthy, Mehrtens, Monson, Murphy, Papovich,
  Persson, Quadri, Rees, Tomczak, Tran, \& Tilvi}]{Straatman2014}
Straatman, C. M.~S., Labb{\'{e}}, I., Spitler, L.~R., {et~al.} 2014, The
  Astrophysical Journal, 783, L14, \dodoi{10.1088/2041-8205/783/1/L14}

\bibitem[{Straatman {et~al.}(2015{\natexlab{b}})Straatman, Labb{\'{e}},
  Spitler, Glazebrook, Tomczak, Allen, Brammer, Cowley, van Dokkum, Kacprzak,
  Kawinwanichakij, Mehrtens, Nanayakkara, Papovich, Persson, Quadri, Rees,
  Tilvi, Tran, \& Whitaker}]{Straatman2015a}
---. 2015{\natexlab{b}}, The Astrophysical Journal, 808, L29,
  \dodoi{10.1088/2041-8205/808/1/L29}

\bibitem[{Tanaka {et~al.}(2019)Tanaka, Valentino, Toft, Onodera, Shimakawa,
  Ceverino, Faisst, Gallazzi, G{\'{o}}mez-Guijarro, Kubo, Magdis, Steinhardt,
  Stockmann, Yabe, \& Zabl}]{Tanaka2019}
Tanaka, M., Valentino, F., Toft, S., {et~al.} 2019, The Astrophysical Journal,
  885, L34, \dodoi{10.3847/2041-8213/ab4ff3}

\bibitem[{Thomas {et~al.}(2013)Thomas, Steele, Maraston, Johansson, Beifiori,
  Pforr, Str{\"{o}}mb{\"{a}}ck, Tremonti, Wake, Bizyaev, Bolton, Brewington,
  Brownstein, Comparat, Kneib, Malanushenko, Malanushenko, Oravetz, Pan,
  Parejko, Schneider, Shelden, Simmons, Snedden, Tanaka, Weaver, \&
  Yan}]{Thomas2013}
Thomas, D., Steele, O., Maraston, C., {et~al.} 2013, Monthly Notices of the
  Royal Astronomical Society, 431, 1383, \dodoi{10.1093/mnras/stt261}

\bibitem[{Toft {et~al.}(2012)Toft, Gallazzi, Zirm, Wold, Zibetti, Grillo, \&
  Man}]{Toft2012}
Toft, S., Gallazzi, A., Zirm, A., {et~al.} 2012, The Astrophysical Journal,
  754, 3, \dodoi{10.1088/0004-637X/754/1/3}

\bibitem[{Toft {et~al.}(2017)Toft, Zabl, Richard, Gallazzi, Zibetti, Prescott,
  Grillo, Man, Lee, G{\'{o}}mez-Guijarro, Stockmann, Magdis, \&
  Steinhardt}]{Toft2017}
Toft, S., Zabl, J., Richard, J., {et~al.} 2017, Nature, 546, 510,
  \dodoi{10.1038/nature22388}

\bibitem[{Tremonti {et~al.}(2004)Tremonti, Heckman, Kauffmann, Brinchmann,
  Charlot, White, Seibert, Peng, Schlegel, Uomoto, Fukugita, \&
  Brinkmann}]{Tremonti2004}
Tremonti, C.~A., Heckman, T.~M., Kauffmann, G., {et~al.} 2004, The
  Astrophysical Journal, 613, 898, \dodoi{10.1086/423264}

\bibitem[{Treu {et~al.}(2010)Treu, Auger, Koopmans, Gavazzi, Marshall, \&
  Bolton}]{Treu2010}
Treu, T., Auger, M.~W., Koopmans, L.~V., {et~al.} 2010, Astrophysical Journal,
  709, 1195, \dodoi{10.1088/0004-637X/709/2/1195}

\bibitem[{Valdes {et~al.}(2004)Valdes, Gupta, Rose, Singh, \&
  Bell}]{Valdes2004}
Valdes, F., Gupta, R., Rose, J.~A., Singh, H.~P., \& Bell, D.~J. 2004, The
  Astrophysical Journal Supplement Series, 152, 251, \dodoi{10.1086/386343}

\bibitem[{Valentino {et~al.}(2020)Valentino, Tanaka, Davidzon, Toft,
  G{\'{o}}mez-Guijarro, Stockmann, Onodera, Brammer, Ceverino, Faisst,
  Gallazzi, Hayward, Ilbert, Kubo, Magdis, Selsing, Shimakawa, Sparre,
  Steinhardt, Yabe, \& Zabl}]{Valentino2020}
Valentino, F., Tanaka, M., Davidzon, I., {et~al.} 2020, The Astrophysical
  Journal, 889, 93, \dodoi{10.3847/1538-4357/ab64dc}

\bibitem[{van~de Sande {et~al.}(2013)van~de Sande, Kriek, Franx, van Dokkum,
  Bezanson, Bouwens, Quadri, Rix, \& Skelton}]{vandeSande2013}
van~de Sande, J., Kriek, M., Franx, M., {et~al.} 2013, The Astrophysical
  Journal, 771, 85, \dodoi{10.1088/0004-637X/771/2/85}

\bibitem[{{Van Der Wel} {et~al.}(2009){Van Der Wel}, Bell, {Van Den Bosch},
  Gallazzi, \& Rix}]{VanDerWel2009}
{Van Der Wel}, A., Bell, E.~F., {Van Den Bosch}, F.~C., Gallazzi, A., \& Rix,
  H.~W. 2009, Astrophysical Journal, 698, 1232,
  \dodoi{10.1088/0004-637X/698/2/1232}

\bibitem[{van~der Wel {et~al.}(2014)van~der Wel, Franx, van Dokkum, Skelton,
  Momcheva, Whitaker, Brammer, Bell, Rix, Wuyts, Ferguson, Holden, Barro,
  Koekemoer, Chang, McGrath, H{\"{a}}ussler, Dekel, Behroozi, Fumagalli, Leja,
  Lundgren, Maseda, Nelson, Wake, Patel, Labb{\'{e}}, Faber, Grogin, \&
  Kocevski}]{vanderWel2014}
van~der Wel, A., Franx, M., van Dokkum, P.~G., {et~al.} 2014, The Astrophysical
  Journal, 788, 28, \dodoi{10.1088/0004-637X/788/1/28}

\bibitem[{van Dokkum {et~al.}(2017)van Dokkum, Conroy, Villaume, Brodie, \&
  Romanowsky}]{vanDokkum2017}
van Dokkum, P., Conroy, C., Villaume, A., Brodie, J., \& Romanowsky, A.~J.
  2017, The Astrophysical Journal, 841, 68, \dodoi{10.3847/1538-4357/aa7135}

\bibitem[{van Dokkum(2008)}]{vanDokkum2008}
van Dokkum, P.~G. 2008, The Astrophysical Journal, 674, 29,
  \dodoi{10.1086/525014}

\bibitem[{van Dokkum {et~al.}(2009{\natexlab{a}})van Dokkum, Kriek, \&
  Franx}]{vanDokkum2009a}
van Dokkum, P.~G., Kriek, M., \& Franx, M. 2009{\natexlab{a}}, Nature, 460,
  717, \dodoi{10.1038/nature08220}

\bibitem[{van Dokkum {et~al.}(2009{\natexlab{b}})van Dokkum, Labb{\'{e}},
  Marchesini, Quadri, Brammer, Whitaker, Kriek, Franx, Rudnick, Illingworth,
  Lee, \& Muzzin}]{vanDokkum2009}
van Dokkum, P.~G., Labb{\'{e}}, I., Marchesini, D., {et~al.}
  2009{\natexlab{b}}, Publications of the Astronomical Society of the Pacific,
  121, 2, \dodoi{10.1086/597138}

\bibitem[{Vazdekis {et~al.}(2010)Vazdekis, S{\'{a}}nchez-Bl{\'{a}}zquez,
  Falc{\'{o}}n-Barroso, Cenarro, Beasley, Cardiel, Gorgas, \&
  Peletier}]{Vazdekis2010}
Vazdekis, A., S{\'{a}}nchez-Bl{\'{a}}zquez, P., Falc{\'{o}}n-Barroso, J.,
  {et~al.} 2010, Monthly Notices of the Royal Astronomical Society, 404, 1639,
  \dodoi{10.1111/j.1365-2966.2010.16407.x}

\bibitem[{Villaume {et~al.}(2017)Villaume, Brodie, Conroy, Romanowsky, \& van
  Dokkum}]{Villaume2017}
Villaume, A., Brodie, J., Conroy, C., Romanowsky, A.~J., \& van Dokkum, P.
  2017, The Astrophysical Journal, 850, L14, \dodoi{10.3847/2041-8213/aa970f}

\bibitem[{Wake {et~al.}(2012)Wake, van Dokkum, \& Franx}]{Wake2012}
Wake, D.~A., van Dokkum, P.~G., \& Franx, M. 2012, The Astrophysical Journal,
  751, L44, \dodoi{10.1088/2041-8205/751/2/L44}

\bibitem[{Wilson {et~al.}(2004)Wilson, Henderson, Herter, Matthews, Skrutskie,
  Adams, Moon, Smith, Gautier, Ressler, Soifer, Lin, Howard, LaMarr, Stolberg,
  \& Zink}]{JWilson2004}
Wilson, J.~C., Henderson, C.~P., Herter, T.~L., {et~al.} 2004, Ground-based
  Instrumentation for Astronomy, 5492, 1295, \dodoi{10.1117/12.550925}

\bibitem[{Wuyts {et~al.}(2009)Wuyts, Franx, Cox, Hernquist, Hopkins, Robertson,
  \& {Van Dokkum}}]{Wuyts2009}
Wuyts, S., Franx, M., Cox, T.~J., {et~al.} 2009, Astrophysical Journal, 696,
  348, \dodoi{10.1088/0004-637X/696/1/348}

\bibitem[{Zahid \& Geller(2017)}]{Zahid2017}
Zahid, H.~J., \& Geller, M.~J. 2017, The Astrophysical Journal, 841, 32,
  \dodoi{10.3847/1538-4357/aa7056}

\bibitem[{Zahid {et~al.}(2016)Zahid, Geller, Fabricant, \& Hwang}]{Zahid2016}
Zahid, H.~J., Geller, M.~J., Fabricant, D.~G., \& Hwang, H.~S. 2016, The
  Astrophysical Journal, 832, 203, \dodoi{10.3847/0004-637x/832/2/203}

\end{thebibliography}

\clearpage
\appendix \label{App}
\renewcommand\thefigure{\thesection.\arabic{figure}}
\renewcommand\thetable{\thesection.\arabic{table}}     

\section{Comparison of MOSFIRE and NIRES spectra for XMM-VID1-2075}\label{App:K2075}
\setcounter{figure}{0}    
\setcounter{table}{0} 

\begin{figure*}[bp]
	\centering{\includegraphics[width=\textwidth,trim=0in 0in 0in 0in, clip=true]{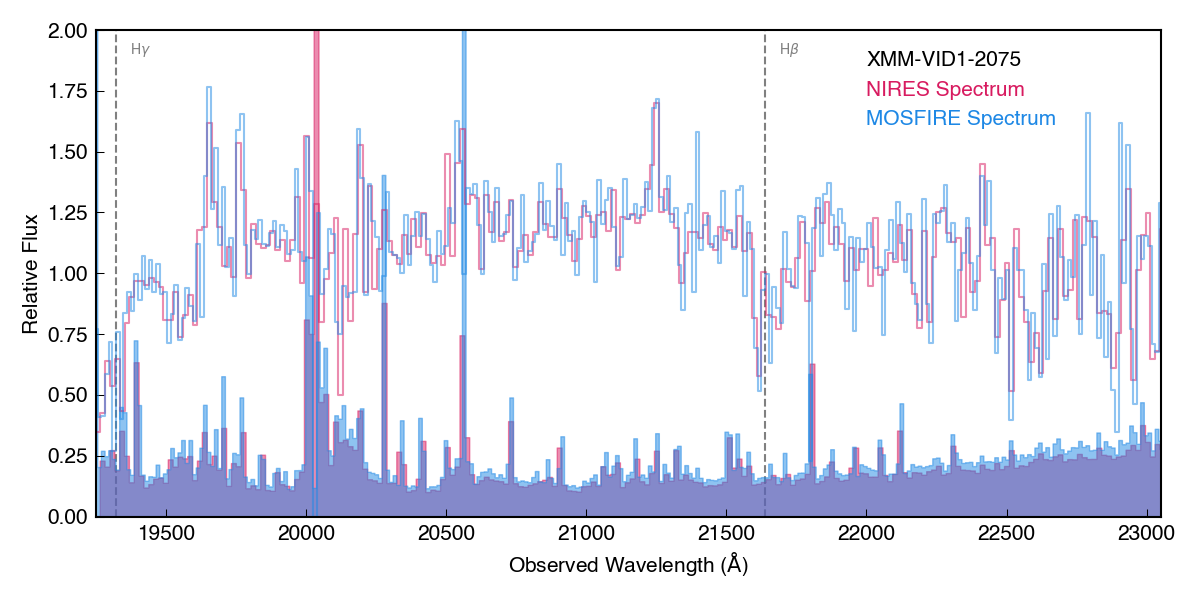}}
	\caption{Observed $K$-band spectra for XMM-VID1-2075 from NIRES (red) and MOSFIRE (blue). Spectra are binned to $\sim$3 ${\rm \AA}$ in the rest-frame. Errors are represented by the shaded regions at the bottom, and the H$\beta$ and $H\gamma$ absorption features at $z=3.45$ are labeled.}
	\label{fig:App_2075}
\end{figure*}

One of the UMGs in this work, XMM-VID1-2075, has both a MOSFIRE $K$-band spectrum and a NIRES spectrum, which also includes both $K$-band and $H$-band data with some signal-to-noise.
The spectra appear quite similar (Figure~\ref{fig:App_2075}).
We fit the $K$-band spectrum from each instrument with the galaxy photometry using FAST++ and compare the results.
The redshifts from the two fits are very similar, with $z_{\rm MOSFIRE}=3.4523$ and $z_{\rm NIRES}=3.4482$, a difference of $\sim 0.1\%$.
In both cases, the best fit indicates a galaxy with \logM$\sim11.5$, $A_V\sim0.3$, and age $\sim 300-500$ Myr.

However, including the NIRES $H$-band data while performing the fit results in a slightly older, less massive, less dusty galaxy (\logM$\sim11.3$, $A_V\sim0$, age $\sim800$ Myr).
When each spectrum is fit with pPXF with a set of inputs and assuming the best-fit redshift to that spectrum, the results are statistically consistent.
In this work we use the values from the fit to the entire $H$- and $K$-band NIRES spectrum, as this provides a greater number of features for determination of the velocity dispersion.


\section{Dependence of Velocity Dispersions on pPXF Inputs}\label{App:pPXF}
\setcounter{figure}{0}    
\setcounter{table}{0} 
Due to the low SNR of our spectra (order $\sim1$/pixel) compared to those pPXF was originally tested on (order $\sim100$/pixel), the resultant velocity dispersions can be sensitive to various parts of the fitting mechanism, including choice of templates, additive polynomial order, and wavelength range, among others.
Extensive tests along these lines have been performed by \citet{vandeSande2013} for a sample of galaxies at $z\sim2$, some of which we reproduce for our sample.

\subsection{Age and Metallicity Template dependence}\label{App:AgeMet}

\begin{figure*}[bp]
	\centering{\includegraphics[width=\textwidth,trim=0in 0in 0in 0in, clip=true]{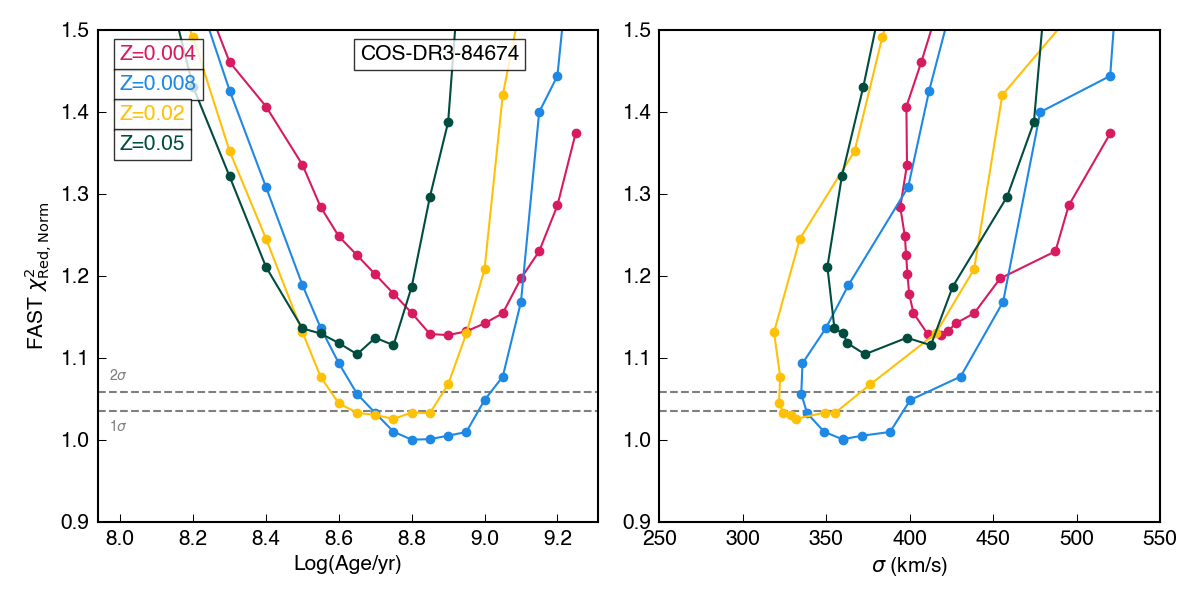}}
	\caption{The reduced $\chi^2$ for COS-DR3-84674 compared to velocity dispersion for BC03 templates demonstrating the age-metallicity degeneracy. The $\chi^2$ values are taken from the FAST++ joint fit to both photometric and spectroscopic data, and are normalized to the lowest value by scaling the input spectral errors. One and two sigma significance given the number of degrees of freedom are indicated by horizontal lines. Each colored line represents templates with a set metallicity, while each point is a different age template.}
	\label{fig:App_dAgeMet}
\end{figure*}

As the spectra herein have low SNR/pixel, slightly different templates can yield similar fits to the spectra alone.
In particular, the degeneracy between age and metallicity can affect line widths and depths in ways that are difficult to disentangle using a low SNR spectrum alone.
These difficulties can be somewhat alleviated by taking into account the broadband photometry of a galaxy.
Similar to \citet{vandeSande2013} and \citet{Hill2016}, we test model dependency in this regime using the BC03 models due to their extended wavelength coverage.
We use FAST++ to fit the spectra and photometry in combination with age and metallicity fixed over a range of values (each combination of $\log$(Age/yr)=[8.0, 8.1, 8.2, 8.3, 8.4, 8.5, 8.55, 8.6, 8.65, 8.7, 8.75, 8.8, 8.85, 8.9, 8.95, 9.0, 9.05, 9.1, 9.15, 9.2, 9.25] and Z=[0.004, 0.008, 0.02 (solar), 0.05]).
We then use pPXF to fit the velocity dispersion of the galaxy using the best-fit template from each combination of age and metallicity, and compare to the reduced $\chi^2$ value from the FAST++ fit.
An example of the results are shown for COS-DR3-84674 in Figure~\ref{fig:App_dAgeMet}.
In all cases, the models show clear minima for each choice of metallicity, though in some cases a particular metallicity is not statistically favored.
The model with the lowest reduced $\chi^2$ was used for this paper and in subsequent tests.
Importantly, this choice is independent of pPXF and therefore also independent of additive polynomial order and spectral wavelength range (see following sections).


\subsection{Dependence on Additive Polynomial Order}\label{App:Poly}

\begin{figure*}[bp]
	\centering{\includegraphics[width=0.5\textwidth,trim=0in 0in 0in 0in, clip=true]{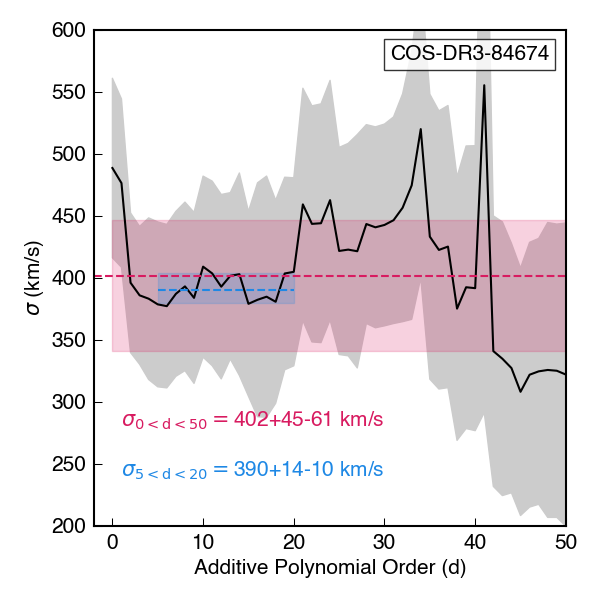}}
	\caption{Velocity dispersion dependence upon choice of additive polynomial order for COS-DR3-84674. Results are calculated for $0\le d \le 50$, with the gray shaded region indicating the uncertainties returned by pPXF for each measurement. The red dashed line indicates the median value across all choices of $d$, shading between the 16th and 84th percentiles.  The blue dashed line is the median value over $5 \le d \le 20$.}
	\label{fig:App_dPoly}
\end{figure*}

The pPXF program allows for addition of a $d$-dimensional Legendre polynomial to a template in order to better match the observed spectrum.
This provides better fits to lower SNR features in the observed spectral line profiles.
A choice of polynomial order which is too low can fail to accurately match the template and observed spectrum, while excessively large order polynomials end up perturbing a template to match observational noise which often yields nonsensical results.
We test the dependence of output velocity dispersion on polynomial order by forcing pPXF to fit the observed spectrum with the single best-fit BC03 template as determined above with order fixed to each $d=[1,2,3,..,50]$.
Example results are shown in Figure~\ref{fig:App_dPoly}.
For the most part, we see the greatest variability in output velocity dispersion at $d>20$, as well as some at $d<5$, while between these values the output velocity dispersion appears generally stable. 


\subsection{Dependence on Spectral Wavelength Range}\label{App:Wave}

\begin{figure*}[tp]
	\centering{\includegraphics[width=0.8\textwidth,trim=0in 0in 0in 0in, clip=true]{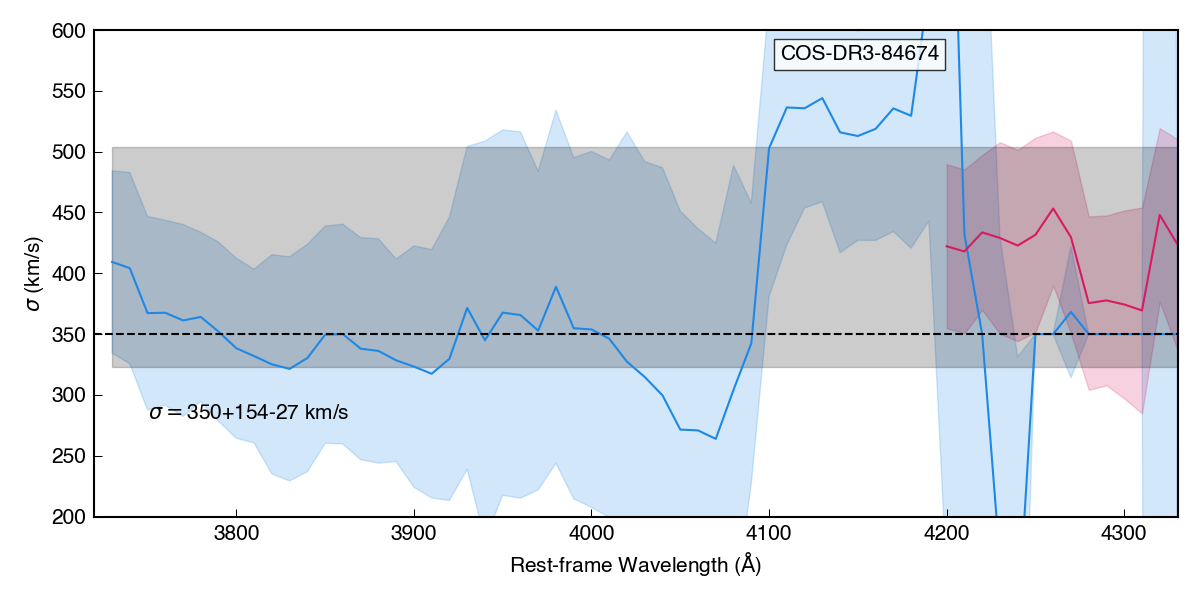}}
	\caption{Velocity dispersion fit as a result of trimming the observed spectra. Blue lines represent the velocity dispersion returned when the spectrum is fit from the rest-frame wavelength on the abscissa to the reddest wavelength available.  Red lines represent the velocity dispersion returned when the spectrum is fit from the bluest wavelength available to the wavelength on the abscissa. Shaded regions show the uncertainties returned by pPXF for each measurement. An additive polynomial of order $d=10$ is used for this example.}
	\label{fig:App_dWave}
\end{figure*}

Velocity dispersion fits are also dependent upon the wavelengths available in the observed spectrum, where the inclusion or exclusion of specific spectral features can alter results.
We refit truncated spectra using a range of starting wavelengths from $3200<\lambda_{\rm rest, blue}/{\rm \AA}<5000$ and ending wavelengths $4200<\lambda_{\rm rest, red}/{\rm \AA}<6000$ and analyze the results (Figure~\ref{fig:App_dWave}).
The most apparent result is that when the spectrum includes the Ca H\&K lines, the velocity dispersion results are significantly more stable.
In many cases, there also appears to be variability with the inclusion or exclusion of H$\delta$.
Notably, we tend not to see much dependence on the H$\beta$ feature, which suggests that there is little line infilling.
Further insights are difficult due to the small sample, low SNR of the spectra, and dependence of results on polynomial order.

Given the strong dependence of the results on the inclusion of Ca H\&K, we also fit the spectra over the narrow range of $3900<\lambda_{\rm rest}/{\rm \AA}<4000$, so as to isolate these features.
However, doing so precludes the use of the higher order polynomials discussed above, as the narrow wavelength range means each order has outsize effects on the result.
Nevertheless, the results of this fit are statistically consistent within $1\sigma$ for 5 of the galaxies.
The remaining 3 (COS-DR1-99209, COS-DR3-111740, and COS-DR3-202019) showed significant deviations at $d=0$ when testing polynomial order above and so this discrepancy is not surprising.


\subsection{Dependence on Template Library}\label{App:Lib}

\begin{figure*}[bp]
	\centering{\includegraphics[width=0.7\textwidth,trim=0in 0in 0in 0in, clip=true]{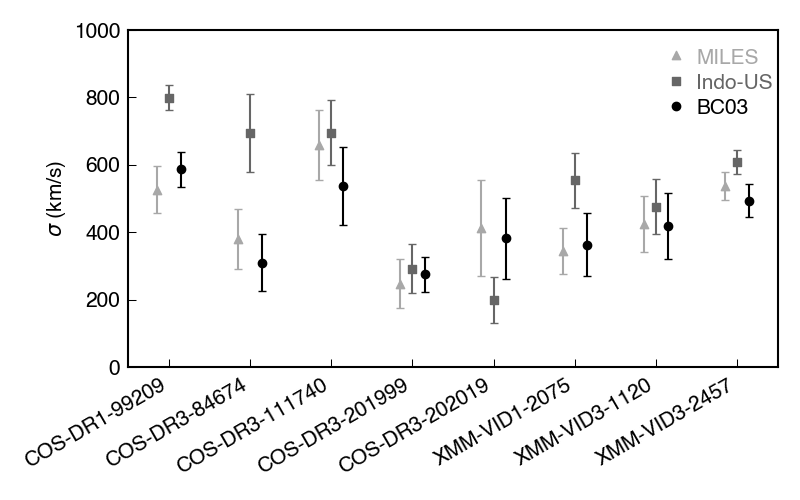}}
	\caption{Velocity dispersion fit as a result of choice of template library with all other parameters fixed. Templates from the MILES library (light gray triangles), Indo-US library (dark gray squares), and Bruzual \& Charlot models (black circles) are shown for each galaxy.}
	\label{fig:App_dLib}
\end{figure*}

In this work we use the BC03 template library due to its longer wavelength coverage, which allows joint fitting with photometry using FAST++.
However, the velocity dispersions from pPXF can be highly dependent upon the availability of templates which are appropriate to the data.
As such, we analyze results from using solely pPXF with three libraries: the BC03 library, the Indo-US stellar library templates \citep{Valdes2004}, and SSPs from the MILES stellar library \citep{Sanchez-Blazquez2006, Vazdekis2010}.
For half of the galaxies all three libraries yield statistically similar results with other parameters fixed (Figure~\ref{fig:App_dLib}), while in the other half the MILES and BC03 outputs are similar and the Indo-US library produces discrepant results.


\subsection{Overall Distribution of Velocity Dispersions}\label{App:Dist}

\begin{figure*}[tp]
	\centering{\includegraphics[width=\textwidth,trim=0in 0in 2in 0in, clip=true]{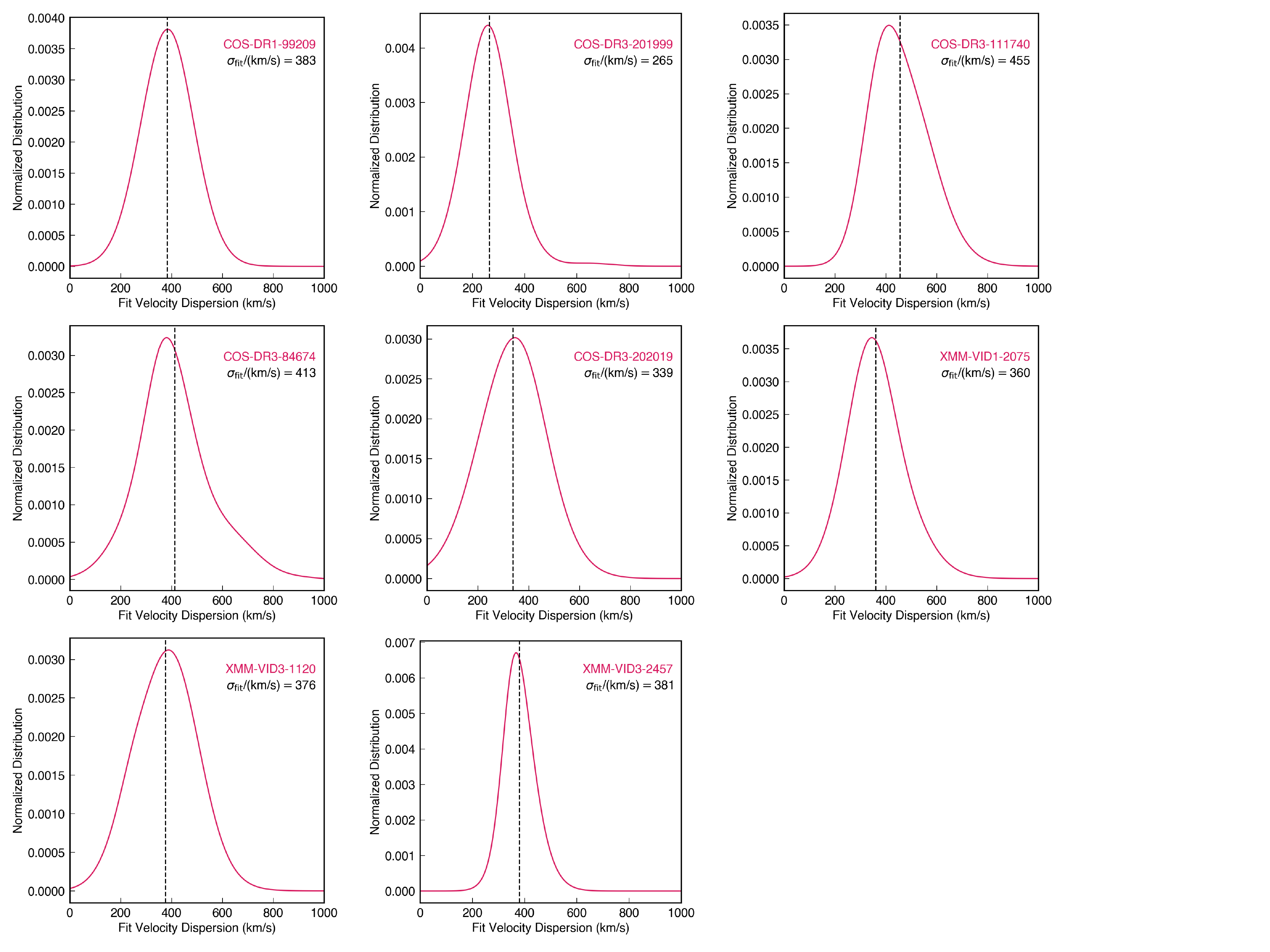}}
	\caption{The distribution of measured velocity dispersions returned by pPXF for each of the MAGAZ3NE objects presented in this work. Here each run of pPXF is convolved with a Gaussian kernel with a width equivalent to the uncertainty reported by pPXF, and additionally weighted by the reduced $\chi^2$ of the fit.  The weighted average is also labeled. This is the value transformed to $\sigma_{\rm e}$ by performing an aperture correction.
	}
	\label{fig:VDdist}
\end{figure*}

As mentioned in the text, we perform a large number of fits with pPXF for each galaxy.
Due to the variety of results and uncertainties associated with any particular fit, we instead use the distribution of results as a whole to determine stellar velocity dispersion for a particular galaxy.
Each fit was convolved with a Gaussian kernel with a standard deviation equal to the reported uncertainty on the velocity dispersion and additionally weighted by the reduced $\chi^2$ of the fit.
The normalized summation of these fits for the 8 MAGAZ3NE UMGs are shown in Figure~\ref{fig:VDdist}.


\end{document}